%% file: main.tex
\documentclass[lettersize,journal]{IEEEtran}
\usepackage{amsmath,amsfonts}
\usepackage{algorithmic}
\usepackage{algorithm}
\usepackage{array}
\usepackage[caption=false,font=normalsize,labelfont=sf,textfont=sf]{subfig}
\usepackage{textcomp}
\usepackage{stfloats}
\usepackage{url}
\usepackage{verbatim}
\usepackage{color,soul}
\usepackage{graphicx}
\usepackage{standalone}
\usepackage{hyperref}
\usepackage{cite}
\usepackage[nolist]{acronym}
\usepackage{mathtools}
\usepackage{amssymb}
\usepackage{stfloats}
\usepackage{tikz}
\usepackage{xcolor}
\usepackage{pgfplots}
\usepackage{standalone}
\usepackage{xcolor}
\usepackage{hyperref}
\usepackage{xfrac}

\acrodefplural{HPA}[HPAs]{high power amplifiers}

\input{acronyms.tex}

\DeclareMathOperator{\MUI}{MUI}
\DeclareMathOperator{\PAPR}{PAPR}
\DeclareMathOperator{\comm}{comm}
\DeclareMathOperator{\ve}{vec}
\DeclareMathOperator{\Real}{Re}
\DeclareMathOperator{\Imag}{Im}
\DeclareMathOperator{\dB}{dB}

\DeclareMathOperator{\bpspHz}{bps/Hz}

\DeclareMathOperator*{\argmax}{arg\,max}
\DeclareMathOperator*{\argmin}{arg\,min}

\usepackage{xcolor,cite,etoolbox}
\makeatletter 
\pretocmd\@bibitem{\color{black}\csname keycolor#1\endcsname}{}{\fail}
\newcommand\citecolor[1]{\@namedef{keycolor#1}{\color{blue}}}
\makeatother
\newtheorem{theorem}{Theorem}

\usepackage{titlesec}

\titlespacing{\subsection}{3pt}{*0.05}{*0.1}
\titlespacing{\subsubsection}{3pt}{*0.05}{*0.1}
\begin{document}
\bstctlcite{MyControl}
\title{\textcolor{black}{DRIP: A Versatile Family of Space-Time ISAC Discrete-time Sequences}}

\author{Dexin Wang, Ahmad Bazzi,  Marwa Chafii 
\thanks{
Dexin Wang is with the NYU Tandon School of Engineering, Brooklyn, 11201, NY, US and the Engineering Division, New York University (NYU) Abu Dhabi, 129188, UAE (email: dw2712@nyu.edu).

Ahmad Bazzi and Marwa Chafii are with Engineering Division, New York University (NYU) Abu Dhabi, 129188, UAE and NYU WIRELESS,
NYU Tandon School of Engineering, Brooklyn, 11201, NY, USA (email: \href{ahmad.bazzi@nyu.edu}{ahmad.bazzi@nyu.edu}, \href{marwa.chafii@nyu.edu}{marwa.chafii@nyu.edu}).}
}

\markboth{Accepted to IEEE Journal on Selected Areas in Information Theory, 2026}%
{Shell \MakeLowercase{\textit{et al.}}: A Sample Article Using IEEEtran.cls for IEEE Journals}

\IEEEpubid{}

\maketitle

\begin{abstract}
The following paper introduces Dual beam-similarity awaRe Integrated sensing and	 communications (ISAC) with controlled Peak-to-average power ratio (DRIP) \textcolor{black}{discrete-time sequences}. 
DRIP is a novel family of space-time ISAC \textcolor{black}{discrete-time sequences} designed for dynamic peak-to-average power ratio (PAPR) adjustment. 
The proposed DRIP \textcolor{black}{sequences} are designed to conform to specified PAPR levels while exhibiting beampattern properties, effectively targeting multiple desired directions and suppressing interference for multi-target sensing applications, while closely resembling radar chirps. 
For communication purposes, the proposed DRIP \textcolor{black}{sequences} aim to minimize multi-user interference across various constellations. 
Addressing the non-convexity of the optimization framework required for generating DRIP \textcolor{black}{sequences}, we introduce a block cyclic coordinate descent algorithm. 
This iterative approach ensures convergence to an optimal ISAC \textcolor{black}{sequence} solution. 
Simulation results validate the DRIP waveforms' superior performance, versatility, and favorable ISAC trade-offs, highlighting their potential in advanced multi-target sensing and communication systems.
\end{abstract}

\vspace{-0.4cm}
\section{Introduction}
\label{sec:introduction}
\input{sections/introduction.tex}

\vspace{-0.5cm}
\section{System Model}
\label{sec:system-model}
\input{sections/system-model.tex}

\vspace{-0.5cm}
\section{Dual Beam-Similarity Aware ISAC Optimization with PAPR Control}
\label{sec:opt-probs}
\input{sections/optimization-problems.tex}

\vspace{-0.5cm}
\section{DRIP \textcolor{black}{Sequences} via Block Cyclic Coordinate Descent}
\label{sec:opt-01}
\input{sections/optimization01.tex}

\input{sections/algo-description-1.tex}

\vspace{-0.5cm}
\section{Convergence Analysis}
\label{sec:convergence-analysis}
\input{sections/convergence-analysis.tex}

\vspace{-0.5cm}
\section{\textcolor{black}{Complexity Analysis}}
\label{sec:complexity-analysis}
\textcolor{black}{\input{actions/complexity-analysis}}

\vspace{-0.5cm}
\section{Simulation Results}
\label{sec:simulations}
\input{sections/simulation-results.tex}

\vspace{-0.5cm}
\section{Conclusions}
\label{sec:conclusions}
\input{sections/conclusions.tex}

\vspace{-0.5cm}
\section*{\textcolor{black}{ACKNOWLEDGMENTS}}
\input{sections/acks.tex}

\vspace{-0.5cm}
\appendices
\label{sec:appendix}
\section{Proof of \textbf{Property 1}}
\label{appendix:proof-QCQP}
\input{sections/proof-qcqp.tex}

\vspace{-0.5cm}
\section{Proof of \textbf{Theorem 1}}
\label{appendix:station-QCQP}
\input{sections/proof-converge-stationary.tex}

\vspace{-0.5cm}
\section{Proof of \textbf{Theorem 2}}
\label{appendix:outer-convergence}
\input{sections/outer-convergence.tex}

\bibliographystyle{IEEEtran}
\bibliography{IEEEabrv,refs}

\vfill

\end{document}

%% file: acronyms.tex
\begin{acronym}
	\acro{AN}{artificial noise}
	\acro{SI}{self interference}
    \acro{MIMO}{multiple-input, multiple-output}
    \acro{AWGN}{additive white Gaussian noise}
    \acro{KPI}{key performance indicator}
      \acro{KPIs}{key performance indicators}  
    \acro{D2D}{device-to-device}
    \acro{RCS}{radar cross-section}
    \acro{SAC}{sensing and communication}
    \acro{ISAC}{integrated sensing and communications}
    \acro{DFRC}{dual-functional radar and communication}
    \acro{SRIP}{Similarity awaRe Integrated sensing and communications with tunable Peak-to-average power ratio}
    \acro{DRIP}{Dual beam-similarity awaRe Integrated sensing and communications with controlled Peak-to-average power ratio}
    \acro{AoA}{angle of arrival}
    \acro{AoAs}{angles of arrival}
    \acro{ToA}{time of arrival}
    \acro{BCCD}{block cyclic coordinate descent}
    \acro{EVM}{error vector magnitude}
    \acro{CPI}{coherent processing interval}
    \acro{eMBB}{enhanced mobile broadband}
    \acro{URLLC}{ultra-reliable low latency communications}
    \acro{mMTC}{massive machine type communications}
	\acro{QCQP}{quadratic constrained quadratic programming}
 \acro{AoI}{age-of-information}
	\acro{BnB}{branch and bound}
	\acro{SER}{symbol error rate}
	\acro{LFM}{linear frequency modulation}
	\acro{BS}{base station}
	\acro{UE}{user equipment}
	\acro{DL}{deep learning}
	\acro{CS}{compressed sensing}
	\acro{PR}{passive radar}
	\acro{LMS}{least mean squares}
	\acro{NLMS}{normalized least mean squares}
	\acro{RIS}{reconfigurable intelligent surface}
	\acro{RISs}{reconfigurable intelligent surfaces}
	\acro{IRS}{intelligent reflecting surface}
	\acro{IRSs}{intelligent reflecting surfaces}
	\acro{LISA}{large intelligent surface/antennas}
	\acro{LISs}{large intelligent surfaces}
	\acro{OFDM}{orthogonal frequency-division multiplexing}
	\acro{EM}{electromagnetic}
	\acro{ISMR}{integrated sidelobe to mainlobe ratio}
	\acro{MSE}{mean squared error}
	\acro{SNR}{signal-to-noise ratio}
	\acro{SRP}{successful recovery probability}
	\acro{CDF}{cumulative distribution function}
	\acro{ULA}{uniform linear array}
	\acro{RSS}{received signal strength}
	\acro{SU}{single user}
	\acro{MU}{multi-user}
	\acro{CSI}{channel state information}
	\acro{OFDM}{orthogonal frequency-division multiplexing}
	\acro{DL}{downlink}
	\acro{i.i.d}{independently and identically distributed}
	\acro{UL}{uplink}
	\acro{LoS}{line of sight}
	\acro{DFT}{discrete Fourier transform}
    \acro{HPA}{high power amplifier}
    \acro{IBO}{input power back-off}
    \acro{MIMO}{multiple-input, multiple-output}
    \acro{PAPR}{peak-to-average power ratio}
    \acro{AWGN}{additive white Gaussian noise}
    \acro{KPI}{key performance indicator}
    \acro{FD}{full duplex}
    \acro{KPIs}{key performance indicators}  
    \acro{D2D}{device-to-device}
    \acro{ISAC}{integrated sensing and communications}
    \acro{DFRC}{dual-functional radar and communication}
    \acro{AoA}{angle of arrival}
    \acro{AoD}{angle of departure}
    \acro{STARS}{simultaneously transmitting and reflecting surface}
    \acro{ToA}{time of arrival}
    \acro{MUI}{multi-user interference}
    \acro{3GPP}{Third Generation Partnership Project}
    \acro{TSG}{Technical Specification Group}
    \acro{SA1}{Service \& System Aspects - Working Group 1}
    \acro{ADAS}{advanced driver assistance systems}
    \acro{QPSK}{quadrature phase shift keying}
    \acro{EVM}{error vector magnitude}
    \acro{CPI}{coherent processing interval}
    \acro{eMBB}{enhanced mobile broadband}
    \acro{URLLC}{ultra reliable and low latency communications}
    \acro{CRB}{Cram\'er-Rao bound}
    \acro{LAA}{lens antenna array}
    \acro{LPI}{low probability of intercept}
    \acro{MI}{mutual information}
    \acro{AI}{artificial intelligence}
    \acro{OTFS}{orthogonal time frequency space}
    \acro{mMTC}{massive machine type communications}
	\acro{QCQP}{quadratic constrained quadratic programming}
	\acro{BnB}{branch and bound}
	\acro{QAM}{quadrature amplitude modulation}
	\acro{PSK}{phase-shift keying}
	\acro{SER}{symbol error rate}
	\acro{BFGS}{Broyden-Fletcher-Goldfarb-Shanno}
	\acro{ADMM}{alternating direction method of multipliers}
	\acro{SQRBS}[SQR BS]{successive QCQP refinement binary search}
	\acro{CCDF}{complementary cumulative distribution function}
	\acro{MISO}{multiple-input, single-output}
	\acro{CSI}{channel state information}
	\acro{LDPC}{low-density parity-check}
	\acro{BCC}{binary convolutional coding}
	\acro{IEEE}{Institute of Electrical and Electronics Engineers}
	\acro{ULA}{uniform linear antenna}
	\acro{B5G}{beyond 5G}
	\acro{MOOP}{multi-objective optimization problem}
	\acro{ML}{maximum likelihood}
	\acro{FML}{fused maximum likelihood}
	\acro{RF}{radio frequency}
	\acro{AM/AM}{amplitude modulation/amplitude modulation}
	\acro{AM/PM}{amplitude modulation/phase modulation}
	\acro{BS}{base station}
	\acro{SCA}{successive convex approximation}
	\acro{MM}{majorization-minimization}
	\acro{LNCA}{$\ell$-norm cyclic algorithm}
	\acro{OCDM}{orthogonal chirp-division multiplexing}
	\acro{TR}{tone reservation}
	\acro{COCS}{consecutive ordered cyclic shifts}
	\acro{LS}{least squares}
	\acro{CVE}{coefficient of variation of envelopes}
	\acro{BSUM}{block successive upper-bound minimization}
	\acro{ICE}{iterative convex enhancement}
	\acro{SOA}{sequential optimization algorithm}
	\acro{BCD}{block coordinate descent}
	\acro{SINR}{signal to interference plus noise ratio}
	\acro{MICF}{modified iterative clipping and filtering}
	\acro{PSL}{peak side-lobe level}
	\acro{SDR}{semi-definite relaxation}
	\acro{KL}{Kullback-Leibler}
	\acro{ADSRP}{alternating direction sequential relaxation programming}
	\acro{MUSIC}{MUltiple SIgnal Classification}
	\acro{EVD}{eigenvalue decomposition}
	\acro{SVD}{singular value decomposition}
	\acro{i.i.d}{independent and identically distributed}
	\acro{UAV}{unmanned aerial vehicle}
	\acro{IoT}{Internet-of-Things}
	\acro{ZF}{zero-forcing}
	\acro{MMSE}{minimum mean square error}
\end{acronym}

%% file: sections/introduction.tex
\IEEEPARstart{T}{\lowercase{he}} sixth generation ($6$G) of wireless communication standard is expected to represent a significant paradigm shift from its predecessors due to the numerous emerging applications \cite{10041914}, such as 
autonomous driving \cite{10283734}, 
\acp{UAV} \cite{8833519}, 
industrial \ac{IoT} \cite{8030483,9078108,9865117},
remote health monitoring, 
digital twins \cite{10489861},
and security \cite{10373185}, with activities in the (sub-)terahertz spectrum \cite{8732419} due to its ability to provide millimeter-level resolution and high-speed communication.
The seamless integration of communication and sensing services, which are provided as distinct functionality in the current wireless standards, is a prerequisite for supporting these services \cite{10044188}. 
As a result, \ac{ISAC} has recently attracted a lot of interest as a $6$G technology enabler \cite{10184987,10188491}, due to its capability of improving spectral and energy efficiencies and to mutually benefit from both services \cite{9737357}.
According to \cite{10124135}, when \ac{SAC} functions are used in tandem, they can benefit from one another mutually, but when they are used in competition, there is an inherent trade-off.
The superiority of an \ac{ISAC} system over a frequency-division \ac{SAC} system in terms of sensing communication rate region is demonstrated using a \ac{MI} framework in \cite{10129042}.
Indeed, \ac{ISAC} systems can simultaneously perform communication and radar tasks by designing and transmitting a single waveform via a fully-shared transceiver. This leads to efficient resource utilization \cite{10124714} with lower system cost and lower power consumption.
Hence, advanced designs for dual-functional waveforms are essential to reconcile the conflicting demands of communication and sensing, while also leveraging integration and coordination benefits.
Even more, additional sophistication arises when it becomes necessary to constrain the \ac{PAPR} of the \ac{ISAC} signal under a specified threshold, due to \ac{RF} limitations. 
Specifically, maintaining low \ac{PAPR} transmissions becomes crucial, when nonlinear \acp{HPA} are part of the \ac{RF} frontend of the transmit chain \cite{meilhac2022digital}.
In principle, low \ac{PAPR} waveforms are desired, as it enables us to tune the \ac{HPA}'s quiescent point as close as possible to the optimal operating point, limiting the impact of clipping.
Certainly, extensive research has been conducted on methods for reducing \ac{PAPR}. For instance, \cite{chafii2016necessary} establishes essential conditions for waveforms with superior \ac{PAPR} compared to \ac{OFDM}, while \cite{wang2021model} introduces \ac{PAPR} reduction techniques tailored for \ac{OFDM}.
In addition, \cite{zhou2020new} employs convex optimization techniques to synthesize sequences that exhibit favorable properties of low \ac{PAPR}, while adhering to spectral mask constraints.
Despite research efforts on \ac{PAPR} reduction and optimization, \ac{ISAC} \textcolor{black}{sequences} should be pragmatically dealt with when considering \ac{PAPR} control due to their dual nature and the conflicting nature of the \ac{SAC} information within the waveform. 

\subsection{Existing Work}
\label{sec:existing}
\textcolor{black}{\input{actions/subareas}}
\textcolor{black}{\input{actions/organize-existing-work}}

\textit{Unlike existing methods, we generate low-\ac{PAPR} space-time \ac{ISAC} \textcolor{black}{sequences} for multi-target sensing waveforms that maintain radar characteristics while enabling effective communication. Additionally, we prove that, thanks to our proposed method, it is always guaranteed to converge to a stable waveform that is optimized for \ac{MU} communications and multi-target sensing while minimizing interferers,} produced through reflection from objects that are not the targets of interest, such as clutter.

\subsection{Contributions and Insights}
\label{subsec:contributions-and-insights}
This work derives a family of space-time \ac{ISAC} \textcolor{black}{sequence} design adhering to desired \ac{PAPR} levels.
In particular, a \ac{DFRC} \ac{BS} intends to communicate with \ac{DL} users, while sensing multiple targets through the received backscattered signal of the same transmitted \ac{ISAC} signal, which is to be designed.
We first formulate an optimization framework that enables us to generate a family of space-time \ac{ISAC} \textcolor{black}{sequences} dedicated for the aforementioned tasks and constraints. 
We then design a \ac{BCCD} method intended to generate such \ac{ISAC} \textcolor{black}{sequences}. 
To that purpose, we have summarized our contributions below.
\begin{itemize}
\item \textbf{Space-time controlled \ac{PAPR} \ac{ISAC}  waveform design}. Due to the random nature of communication waveforms, which typically results in degraded sensing performance due to several factors, such as a high \ac{PAPR} and random autocorrelation properties, a fundamental contribution of this work is to design space-time \ac{ISAC} \textcolor{black}{sequence} conveying random communication information, with a closeness to a given radar waveform, hence preserving the radar autocorrelation features. Moreover, the proposed \ac{ISAC} \textcolor{black}{sequences} are capable to simultaneously steer/beam towards intended directions, while rejecting interfering directions, deeming them suitable for \textit{multi-target sensing} applications. In addition, the \ac{PAPR} of the proposed \ac{ISAC} \textcolor{black}{sequences} can be controlled to generate low-\ac{PAPR} waveforms, which is yet another desirable feature in future systems. An optimization framework is derived to accommodate the aforementioned \ac{ISAC} features.
\item \textbf{Optimization method to generate space-time \ac{ISAC} \textcolor{black}{sequences}.} As the proposed optimization framework is non-convex by nature, we propose an iterative \ac{BCCD} algorithm that can be readily used to generate the described space-time \ac{ISAC} \textcolor{black}{sequences}.
\textcolor{black}{\input{actions/baseband-optimization.tex}}
	\item \textbf{Convergence Analysis}. We provide a rigorous convergence analysis for the proposed \ac{BCCD} algorithm for space-time \ac{ISAC} \textcolor{black}{sequence} generation. Specifically, by adopting tools from numerical optimization, it is revealed that the proposed space-time \ac{ISAC} \textcolor{black}{sequence} design method is always guaranteed to converge to a stable waveform, despite the stringent sensing and \ac{PAPR} constraints included in the proposed optimization problem.
	\item \textbf{Extensive simulation results}. Extensive simulation results are provided to compare the performance of the proposed \ac{ISAC} \textcolor{black}{sequence} design method with various benchmark schemes, in terms of average sum rate while trading off the similarity rate, which demonstrates the great potential of our proposed method for \ac{ISAC} systems under a controllable \ac{PAPR} constraint. Moreover, we also reveal convergence performances of the various \ac{ISAC} metrics included, i.e. the average radar \ac{SINR} and \ac{MUI}, in addition to the \ac{PAPR} statistics and the various \ac{ISAC} tradeoffs that can be expected by adopting the proposed framework.
\end{itemize}
Furthermore, we unveil some important insights, i.e.
\begin{itemize}
	\item Simulations reveal that the proposed \ac{BCCD} algorithm requires few iterations to converge towards a \textcolor{black}{stable sequence}.
	\item A similarity-rate tradeoff exists for \ac{DRIP}-type waveforms. In particular, for a very demanding stringent similarity constraint, an average sum rate of about $1.25\bpspHz$ is achieved, whereby relaxing similarity can boost the average sum rate to as high as $6\bpspHz$. 
	\item The analysis of \ac{DRIP} beampatterns\footnote{The resulting beampatterns obtained by the \ac{DRIP} waveforms.} reveals that  increasing similarity to a radar chirp improves sidelobe performance but results in ambiguous beam peaks. When the radar chirp similarity decreases, \ac{DRIP} waveforms produce sharper peaks directed at targets, enhancing beamforming applications. Notably, targets further away achieve clearer peaks even at lower similarity rate values. For instance, one scenario shows distinct peaks of $48 \dB$ and $43 \dB$, while another scenario shows an ambiguous plateau around $37 \dB$ as soon as targets become closer, which is related to the spatial resolution of \ac{DRIP}.
\end{itemize}


\textbf{Notation}: Upper-case and lower-case boldface letters denote matrices and vectors, respectively. $(.)^T$, $(.)^*$ and $(.)^H$ represent the transpose, the conjugate and the transpose-conjugate operators, respectively. The statistical expectation is denoted as $\mathbb{E}\lbrace \rbrace$. For any complex number $z \in \mathbb{C}$, the magnitude is denoted as $\vert z \vert$, and its angle is $ \angle z$. The $\ell_2$ norm of a vector $\pmb{x}$ is denoted as $\Vert \pmb{x} \Vert$. The matrix $\pmb{I}_N$ is the identity matrix of size $N \times N$. The zero-vector is $\pmb{0}$. For matrix indexing, the $(i,j)^{th}$ entry of matrix $\pmb{A}$ is denoted by $[\pmb{A}]_{i,j}$ and its $j^{th}$ column is denoted as $[\pmb{A}]_{:,j}$. The $[x]^+$ operator returns the maximum between $x$ and $0$. A positive semi-definite matrix is denoted as $\pmb{A} \succeq \pmb{0}$ and a vector $\pmb{x}$ with all non-negative entries is denoted as $\pmb{x} \succeq \pmb{0}$. The all-ones vector of appropriate dimensions is denoted by $\pmb{1}$. The vectorization operator is denoted as $\ve$. \textcolor{black}{We use $\mathcal{O}()$ for \textit{big-Oh}.} 
\textcolor{black}{\input{actions/define-phi-Phi}}
\textcolor{black}{$\pmb{P}_{\pmb{B}}$ is the projector matrix onto the column space of $\pmb{B}$.}

%% file: actions/subareas.tex
Recent contributions to \ac{ISAC} waveform design can generally be categorized into three distinct streams: performance metric-driven optimization, modulation-specific waveforms, and metasurface-assisted \ac{ISAC}.

%% file: actions/organize-existing-work.tex
\paragraph{Performance-metric optimization}
In \cite{10521742}, the authors look into joint scheduling and beamforming for \ac{ISAC}-enabled systems under \ac{URLLC} requirements, whereby a joint scheduling and beamforming framework is proposed for joint data transmission and environment sensing. In \cite{10521567}, the authors first derive the expression for \ac{CRB} of \ac{AoA} and distance estimate for sensing multiple wideband sources, and then minimize that while guaranteeing a communication quality-of-service for each downlink user assuming a near-field channel model. Moreover, in \cite{10507060}, the authors design a method for designing waveforms in cooperative joint radar-communications systems, which takes into account the \ac{AoI}. Furthermore, the work in \cite{10445319} studies \ac{ISAC} waveform design for communication-centric systems targeted to maximize energy efficiency, under \ac{CRB} guarantees for target detection and \ac{SINR} guarantees for communication.

\paragraph{Modulation-specific design}
\cite{10522592} uses \ac{OFDM} for \ac{ISAC} waveform design to suppress multi-tone narrowband interference via an optimization framework that trades off between data rate for communications and peak sidelobe levels for sensing. The work in \cite{10498088} proposes a beamspace waveform design and beam selection problem for \ac{ISAC} systems equipped with \acp{LAA}, whereby the goal is to consume minimum system power, under \ac{SAC} constraints. Furthermore, a \ac{LPI} \ac{ISAC} waveform design was proposed in \cite{10304613} for radar systems to jointly attain a low \ac{LPI} for radar functionality and the ability to maintain communication with radar nodes. The work in \cite{10516613} designs a receiver window for \ac{OTFS} signaling in \ac{ISAC} systems and develops a barycenter calibration method that considers high-order spectra of channels in the delay-doppler domain.

\paragraph{Metasurface-assisted \ac{ISAC}}
The work in \cite{10478709} considers \ac{ISAC} waveform design with the aid of \ac{RIS}. In particular, the authors in \cite{10478709} consider secure waveform design with the help of \ac{RIS}, where an \ac{ISAC} base station communicates with multiple users in the presence of an eavesdropper. The work in \cite{10050406} uses \ac{STARS}, whereby sensors are placed at \ac{STARS} to address the significant path loss of sensing. In particular, the $2$D \ac{CRB} for joint \ac{AoA} and \ac{AoD} was minimized under a communication requirement. 

%% file: actions/baseband-optimization.tex
\textcolor{black}{The paper focuses on discrete-time signal processing for \ac{ISAC} rather than continuous-time signaling. All optimizations and designs are therefore done in baseband.}

%% file: actions/define-phi-Phi.tex
We define complex-to-real operators of vectors as $\phi(\pmb{x}) = 
	 \begin{bmatrix}
	\Real^T(\pmb{x}) & \Imag^T(\pmb{x}) 	
	\end{bmatrix}^T$ and that of matrices as 
	$
	\phi(\boldsymbol{A})
	 = 
	 \begin{bmatrix}
	\Real(\pmb{A}) & -\Imag(\pmb{A})  \\
	\Imag(\pmb{A}) & \Real(\pmb{A})  	
	\end{bmatrix} $.

%% file: sections/system-model.tex
\begin{figure}[!t]
\centering
\includegraphics[width=3in]{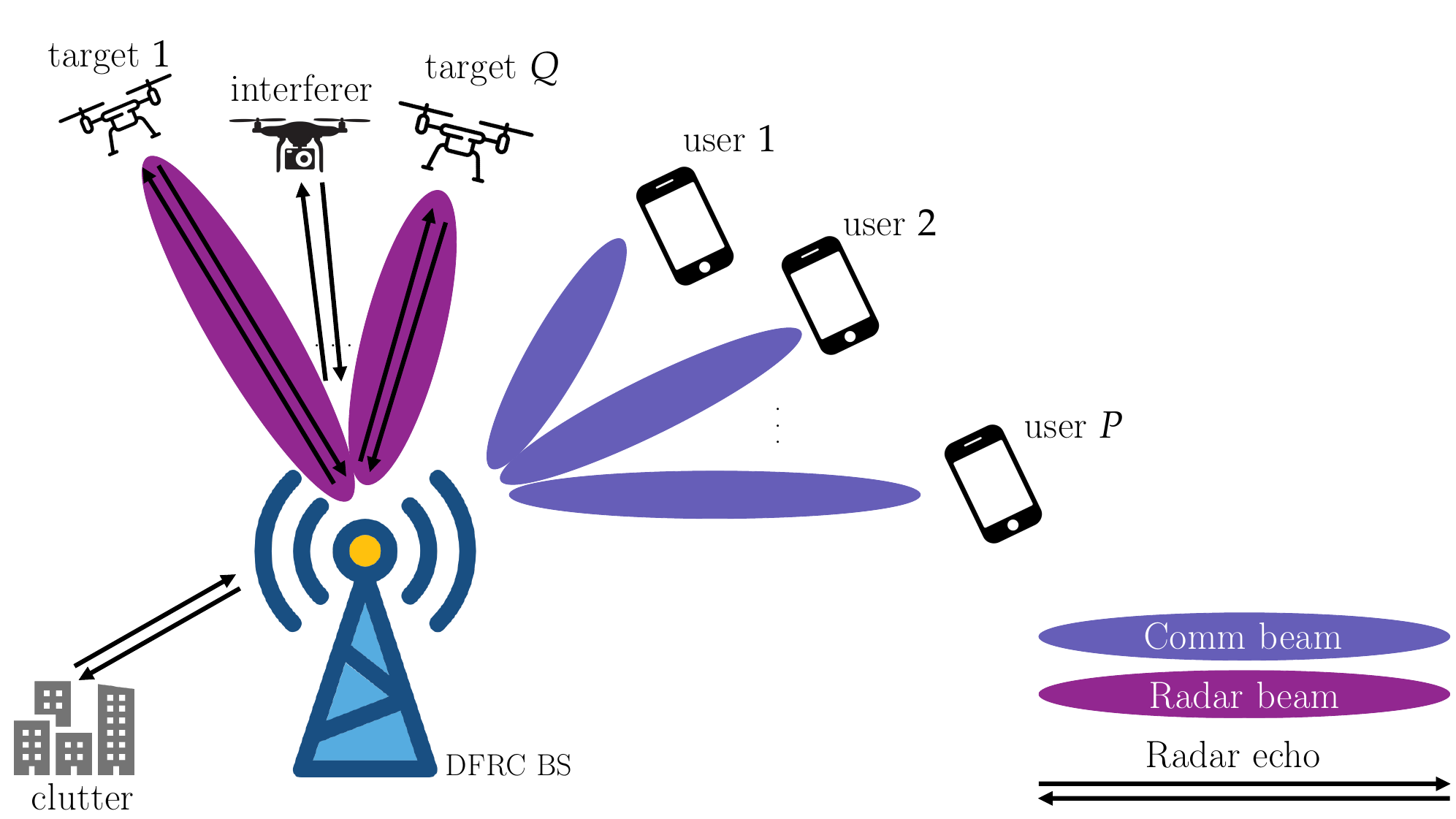}
\caption{An \ac{ISAC} scenario composed of $P$ users for \ac{MU}-communications and $Q$ multiple targets in the presence of $I$ interferers.}
\label{fig:fig_1}
\end{figure}
Let us assume that a \ac{DFRC} \ac{BS}, supporting a co-located mono-static unit of $N_T$ transmit antennas and $N_R$ receive antennas, is to support \ac{MU} communications toward \textcolor{black}{$P$ \ac{DL} communication} single-antenna users, while sensing $Q$ targets\footnote{The targets do not have any transceiver capability.} in the vicinity, in the presence of $I$ interfering targets, which can also be classified as clutter. Moreover, the radar targets of interest are assumed to be positioned at $\theta_1 \ldots \theta_Q$, representing the \ac{AoAs} relative to the \ac{DFRC} \ac{BS}, whereas the interfering targets are assumed to be located at $\bar{\theta}_1 \ldots \bar{\theta}_I$. An illustration is given in Fig. \ref{fig:fig_1}.

\subsection{Multi-user Communication Model}
\label{subsec:comm-sys-model}
\textcolor{black}{\input{actions/modulation-scheme.tex}}

\subsection{Radar Backscattered Model}
\label{subsec:radar-system-model}
Since we assume a mono-static configuration at the \ac{DFRC} \ac{BS}, the \ac{AoD} and 
\ac{AoA} of the different backscattering components are essentially identical. To this extent, we can write
\begin{equation}
	\label{eq:yr1}
\begin{split}
	\pmb{Y}_r &= 
	\sum\nolimits_{q=1}^Q 
	\gamma_q \pmb{\Pi}(\theta_q)\pmb{X} 
	+ \sum\nolimits_{i=1}^I 
	\bar{\gamma}_i \pmb{\Pi}(\bar{\theta}_i)\pmb{X} 
	+ \pmb{Z}_r,
\end{split}
\end{equation}
where $\pmb{Y}_r \in \mathbb{C}^{N_R \times L}$ is the received backscattered echo.
In addition, $\gamma_q$ represents the two-way complex channel coefficient between the \ac{DFRC} \ac{BS} and the $q^{th}$ radar target of interest, which is proportional to the \ac{RCS} of the targets, \textcolor{black}{\input{actions/add-delay-doppler}}In addition, $\bar{\gamma}_i$ is the two-way channel coefficient between the \ac{DFRC} and the $i^{th}$ interferer. Note that the clutter can be accommodated within the interfering terms.
\textcolor{black}{\input{actions/angles-known.tex}}
\textcolor{black}{Referring to \eqref{eq:yr1}}, we have that $\pmb{\Pi}(\theta) = \pmb{a}_{N_R}(\theta) \pmb{a}_{N_T}^T(\theta)$, where $\pmb{a}_{N_T}(\theta) \in \mathbb{C}^{N_T \times 1}$ and $\pmb{a}_{N_R}(\theta) \in \mathbb{C}^{N_R \times 1}$ denote the transmitting and receiving array steering vectors at angle $\theta$, respectively. For example, if the antenna array follows a \ac{ULA} setting, then the steering vectors of transmit and receive arrays can be expressed as \textcolor{black}{$\pmb{a}_{N_T}
	(\theta)
	=
	\begin{bmatrix}
		1 & e^{j \frac{2\pi}{\lambda} d_T \sin(\theta)} & \ldots & e^{j \frac{2\pi}{\lambda} d_T(N_T-1) \sin(\theta)}
	\end{bmatrix}^T$ and $\pmb{a}_{N_R}
	(\theta)
	=
	\begin{bmatrix}
		1 & e^{j \frac{2\pi}{\lambda} d_R \sin(\theta)} & \ldots & e^{j \frac{2\pi}{\lambda} d_R(N_R-1) \sin(\theta)}
	\end{bmatrix}^T$, respectively, where} $\lambda$ is the wavelength; whereas, the inter-element spacing of transmit and receive arrays are denoted as $d_T$ and $d_R$, respectively.
\textcolor{black}{\input{actions/radar-noise-variance}}
In contrast to our previous approach \cite{10061453}, \textit{we hereby aim to design a controlled \ac{PAPR} space-time \ac{ISAC} \textcolor{black}{sequence}} $\pmb{X} \in \mathbb{C}^{N_T \times L}$, under joint multi-target radar beamforming and similarity with respect to a chirp with desirable auto-correlation properties. Therefore, this article offers a sophisticated capability of focusing the \ac{PAPR}-controlled chirp energy towards multiple targets (instead of broadcasting it), in addition to rejecting interfering sources, while performing \ac{MU}-communications. As shown in the next subsection, the same waveform $\pmb{X}$ is used for \ac{MU} communication tasks.

\subsection{\ac{ISAC} Metrics}
\label{subsec:metrics}
\paragraph{Radar chirp similarity}The first metric is the similarity constraint with a desired radar chirping signal. 
\textcolor{black}{\input{actions/KL-divergence}}

\textcolor{black}{\input{actions/delaydoppler}}
\paragraph{\textcolor{black}{Probability of Detection and Radar \ac{SINR}}}Moreover, for space-time beamforming, we define $\pmb{u}_{q,\ell}$ as the beamforming vector intended for the $q^{th}$ target and the $\ell^{th}$ time instance, i.e. \textcolor{black}{the received signal in \eqref{eq:yr1}} is combined as
\begin{equation}
	\label{eq:yr}
\begin{split}
	\pmb{u}_{q,\ell}^H [\pmb{Y}_r]_{:,\ell} &= 
	\gamma_{q} \pmb{u}_{q,\ell}^H\pmb{\Pi}(\theta_{q})\pmb{X}_{:,\ell}
	+
	\sum\nolimits_{q' \neq q}
	\gamma_{q'} \pmb{u}_{q,\ell}^H\pmb{\Pi}(\theta_{q'})\pmb{X}_{:,\ell} 
	\\ &
	+ \sum\nolimits_{i=1}^I 
	\bar{\gamma}_i \pmb{u}_{q,\ell}^H\pmb{\Pi}(\theta_i)\pmb{X}_{:,\ell} 
	+ \pmb{u}_{q,\ell}^H[\pmb{Z}_r]_{:,\ell},
\end{split}
\end{equation}
The first term contains the target of interest, i.e. $q^{th}$ target. The second term contains inter-target interference and the third term contains interference from interfering targets/clutter and the last term is radar noise. 
To this end, we can define the radar \ac{SINR} associated with the $q^{th}$ target \textcolor{black}{as}
\textcolor{black}{\input{actions/gq-replace.tex}}

\textcolor{black}{\input{actions/PD.tex}}

%

\paragraph{Peak-to-average power ratio}
\label{PAPR-paragraph}
The \ac{PAPR} is a waveform measure that reflects the waveform's peak values to its average power.
For instance, a constant magnitude waveform exhibits a unit $\PAPR$, which is the minimal \ac{PAPR} a waveform can have.
\textcolor{black}{\input{actions/papr-explained}}
\paragraph{\ac{MU} interference for communications}Based on \eqref{eq:y=Hxpz}, we can re-write the received signal just by adding and subtracting the desired signal carrying information symbols $\pmb{S} \in \mathbb{C}^{P \times L}$ as
\begin{equation}
	\label{eq:received-signal-2}
	\pmb{Y}_c = \pmb{S} + \pmb{HX} - \pmb{S} + \pmb{Z}_c.
\end{equation} 
Under a fixed transmit constellation energy, minimizing the total energy within the \ac{MU} interference contributes to maximizing the achievable sum-rate of all communication users \cite{mohammed2013per}. In essence, the total \ac{MU} interference energy is given as
\begin{equation}
\label{eq:EMUI}
	E_{\MUI} = \Vert \pmb{HX} - \pmb{S} \Vert_F^2.
\end{equation}
In the next section, we introduce an optimization framework dedicated to generate space-time \ac{ISAC} \textcolor{black}{sequences} that are capable of adhering to a given \ac{PAPR}, while performing communication and space-time sensing tasks.
\textcolor{black}{\input{actions/on-S-X}}

\vspace{-0.1cm}

\textcolor{black}{\input{actions/integrated-process}}

%% file: actions/modulation-scheme.tex
Consider a wireless transmission in the \ac{DL} communication sense with multiple propagation paths between user $p$ and the \ac{DFRC} \ac{BS} with $N_T$ antennas. The received analog signal $y_p(t)$ at user $p$ is the convolution of the transmitted signal $x(t)$ and the channel impulse response, which consists of $K$ paths, each with a pathloss attenuation $\alpha_{p,k,n}$ (which can also contain angle-of-departure and angle-of-arrival information) and time delay $\tau_{p,k,n}$ as 
\begin{equation}
\label{eq:analog1}
	y_{p}(t) = \sum\nolimits_{n=1}^{N_T}\sum\nolimits_{k=1}^{K} \alpha_{p,k,n} x_n(t - \tau_{p,k,n}) + z_p(t),
\end{equation}
$\forall p = 1 \ldots P$, 
where $z_p(t)$ is \ac{AWGN} noise at user $p$ following $z_p(t) \sim \mathcal{CN}(0,\sigma_c^2 )$, and $\sigma_c^2$ represents the communication noise variance power.
We assume that the narrowband single-carrier system occupies bandwidth $B$ and is centered at the carrier $f_c$, under the assumption that $B \ll B_c$, where $B_c$ is the coherence bandwidth of the channel. In other words, the symbol duration $T_s \approx 1/B$ is much larger than the maximum delay spread.
Under this assumption, the signal envelope $x_n(t)$ changes very slowly compared to the delays, i.e. slow-fading block Rayleigh fading. Therefore, we can approximate
\begin{equation}
\label{eq:analog2}
	x_n(t - \tau_{p,k,n}) \approx x_n(t) e^{-j 2\pi f_c  \tau_{p,k,n}}.
\end{equation}
Substituting \eqref{eq:analog2} in \eqref{eq:analog1}, we get 
\begin{equation}
	y_p(t) = \sum\nolimits_{n=1}^{N_T} h_{p,n} x_n(t) + z_p(t)
\end{equation}
where $h_{p,n} = \sum_{k=1}^{K}\alpha_{p,k,n} e^{-j 2\pi f_c  \tau_{p,k,n}}$.
Sampling the signal to digital by sampling at intervals $t = \ell T_s$, where $\ell$ is the sample index, one gets
\begin{equation}
	y_p[\ell] =  \sum\nolimits_{n=1}^{N_T} h_{p,n} x_n[\ell] + z_p[\ell],
\end{equation}
$\forall \ell = 1 \ldots L$, where $L$ is the total number of samples collected over time. 
Stacking in matrix form, we get
\begin{equation}
	\label{eq:y=Hxpz}
	\pmb{Y}_c = \pmb{H}\pmb{X} + \pmb{Z}_c,
\end{equation}
where $[\pmb{Y}_c]_{p,\ell} = y_p[\ell] $, and $[\pmb{H}]_{p,n}  = h_{p,n}$ and for simplicity we denote $\pmb{H}= \begin{bmatrix} \pmb{h}_1 & \pmb{h}_2 & \ldots & \pmb{h}_P \end{bmatrix}^T \in \mathbb{C}^{P \times N_T}$. In addition, $[\pmb{X}]_{n,\ell} = x_n[\ell]$.  Finally, $[\pmb{Z}_c]_{p,\ell} = z_p[\ell]$. The \ac{DFRC} \ac{BS} aims to communicate a constellation $\pmb{S} \in \mathbb{C}^{P \times L}$, e.g. \ac{QAM} or \ac{PSK}, towards the $P$ users. Therefore, the \ac{DFRC} \ac{BS} is tasked to design the discrete sequence $\pmb{X}$ to convey the constellation $\pmb{S}$.
\input{actions/on-H.tex}

%% file: actions/on-H.tex
The channel matrix $\pmb{H}$ is assumed to be perfectly known at the \ac{DFRC} \ac{BS}.
The channel between the \ac{DFRC} \ac{BS} and the $p^{th}$ user is modeled as 
\begin{equation}
	\pmb{h}_p = \bar{\pmb{h}}_p
	+ \sum\nolimits_{q=1}^Q \pmb{h}_{p,q}
	+\sum\nolimits_{i=1}^I \bar{\pmb{h}}_{p,i},
\end{equation}
where $\bar{\pmb{h}}_p$ is the channel between the \ac{DFRC} \ac{BS} and the $p^{th}$ user,
$\pmb{h}_{p,q}$ is the channel between the \ac{DFRC} \ac{BS} towards the $q^{th}$ target, then towards the $p^{th}$ user, and 
$ \bar{\pmb{h}}_{p,i}$ is the channel between the \ac{DFRC} \ac{BS} towards the $qi^{th}$ interferer, then towards the $p^{th}$ user.
All three components follow the Rician model, for instance 
\begin{equation}
	\boldsymbol{h}_{p,q}=\sqrt{\frac{\kappa_{p,q}}{\kappa_{p,q}+1}} \boldsymbol{h}_{p,q}^{\mathrm{LoS}}+\sqrt{\frac{1}{\kappa_{p,q}+1}} \boldsymbol{h}_{p,q}^{\mathrm{NLoS}},
\end{equation}
where $ \boldsymbol{h}_{p,q}^{\mathrm{LoS}}$ is the \ac{LoS} component. Note that $ \boldsymbol{h}_{p,q}^{\mathrm{LoS}}$ implicitly contains the \ac{RCS} and the angles of departure/arrival. For the non-\ac{LoS}, the $\boldsymbol{h}_{p,q}^{\mathrm{NLoS}}$ is modeled as \ac{i.i.d} complex Gaussian random variables, and $\kappa_{p,q}$ represents the Rician K-factor  \cite{10373185,11313766}. The same model has been adopted on $\bar{\pmb{h}}_p$ and $\bar{\pmb{h}}_{p,i}$.

%% file: actions/add-delay-doppler.tex
which includes delay-Doppler information, as well \cite{11275173}.

%% file: actions/angles-known.tex
The angles of targets $\theta_1 \ldots \theta_Q$ and interferers $\bar{\theta}_1 \ldots \bar{\theta}_I$ are known perfectly by the \ac{DFRC} \ac{BS}.

%% file: actions/radar-noise-variance.tex
The noise of the radar sub-system follows an \ac{i.i.d} Gaussian distribution as $\ve(\pmb{Z}_r) \sim \mathcal{CN}(0,\sigma_r^2 \pmb{I}_{N_RL})$.

%% file: actions/KL-divergence.tex
Denoting the desired radar signal as $\pmb{X}_0 \in \mathbb{C}^{N_T \times L}$, we define two probability distributions for the received signal under Gaussian noise, namely $p(\pmb{Z})$, which is the distribution when $\pmb{X}$ is transmitted, i.e. $\pmb{Z}\sim \mathcal{CN}(\pmb{X}, \sigma_r^2\pmb{I})$ and $q(\pmb{Z})$ is that when $\pmb{X}_0$ is transmitted, i.e. $\pmb{Z}\sim \mathcal{CN}(\pmb{X}_0, \sigma_r^2\pmb{I})$. The \ac{KL}-divergence 
\begin{equation}
	D_{\mathrm{KL}}(P \| Q) = \int p(\pmb{Z}) \ln \left( \frac{p(\pmb{Z})}{q(\pmb{Z})} \right) d\pmb{Z} = \mathbb{E}_P \left[ \ln p(\pmb{Z}) - \ln q(\pmb{Z}) \right],
\end{equation}
which under the assumed distributions can be expressed as 
\begin{equation}
	D_{\mathrm{KL}}(P \| Q) = \frac{1}{\sigma_r^2} \Vert\pmb{X} - \pmb{X}_0\Vert^2\propto \|\pmb{x} - \pmb{x}_0\|^2,
\end{equation}
where $\pmb{x} = \ve(\pmb{X})$ and $\pmb{x}_0 = \ve(\pmb{X}_0)$. 
Hence, considering the \ac{KL}-divergence for optimization naturally involves the similarity constraint $\Vert\pmb{X} - \pmb{X}_0\Vert^2$ for minimization.
The similarity constraint can be formed as a sphere centered at the reference radar signal $\pmb{X}_0$ with a radius $\epsilon$, namely $\pmb{x} \in \mathcal{B}_{\epsilon}(\pmb{x}_0)$, where $\mathcal{B}_{\epsilon}(\pmb{x}_0) = \lbrace \pmb{x} , \Vert \pmb{x} - \pmb{x}_0 \Vert^2 \leq \epsilon^2 \rbrace$.

%% file: actions/delaydoppler.tex
A similarity constraint is imposed on the transmitted sequence in order to adapt some correlation properties, including range-Doppler resolution and the peak sidelobe level for the cross-correlation function of the ambiguity function \cite{6850145}.

%% file: actions/gq-replace.tex
\begin{equation}
\label{eq:radar-sinr}
\begin{split}
	g_q 
	&=
	\frac{\sigma_q^2 \sum\nolimits_{\ell = 1}^L \Vert  \pmb{u}_{q,\ell}^H \pmb{\Pi}(\theta_q) \pmb{X}_{:,\ell} \Vert^2 }
	{\sum\nolimits_{\ell = 1}^L
	P_{\mathrm{I+N}}^{(\ell)}},
\end{split}
\end{equation}
and
\begin{equation}
\begin{split}
	P_{\mathrm{I+N}}^{(\ell)}
&=
	\Big(
	 \sum\nolimits_{\substack{ q' \neq q}}^Q
	\sigma_{q'}^2 \Vert  \pmb{u}_{q,\ell}^H \pmb{\Pi}(\theta_{q'}) \pmb{X}_{:,\ell} \Vert^2 \\&
	+
	\sum\nolimits_{i = 1}^I
	\bar{\sigma}_{i}^2
	 \Vert  \pmb{u}_{q,\ell}^H \pmb{\Pi}(\bar{\theta}_{i}) \pmb{X}_{:,\ell} \Vert^2
	 +
	 \sigma_r^2 \Vert \pmb{u}_{q,\ell} \Vert^2
	\Big),
\end{split}
\end{equation}
is the power of the interference-plus-noise components of the $\ell^{th}$ sample after receive beamforming, with $\bar{\sigma}_i^2 = \mathbb{E}( \vert \bar{\gamma}_i \vert^2 )$ and $\sigma_q^2 = \mathbb{E}( \vert \gamma_q \vert^2 )$.

%% file: actions/PD.tex
The likelihood ratio test corresponding to the $q^{th}$ hypothesis test, intended to detect the $q^{th}$ target, we can the sensing detector which is given as $g_q \gtrless_{\mathcal{H}_0}^{\mathcal{H}_1} \mathcal{T}$, 
where $\mathcal{T}$ is the detection threshold. 
According to the detector and under a fixed false-alarm probability, denoted as $\operatorname{P}_{\mathrm{FA}}$, the probability of detection to detect the $q^{th}$ target $\operatorname{P}_{\mathrm{D}}^{(q)}$ can be derived as
\begin{equation}
\label{eq:PD}
	\operatorname{P}_{\mathrm{D}}^{(q)}=Q\left(\sqrt{2  g_q }, \sqrt{F_{\chi_{2}^2}^{-1}\left(1-\operatorname{P}_{\mathrm{FA}}\right)}\right),
\end{equation}
where $Q$ refers to the Marcum Q function, $F_{\chi_{2}^2}^{-1}$ represents the inverse cumulative distribution function (CDF) of the chi-square distribution with order $1$, and the $g_q$ is given by \eqref{eq:radar-sinr}.
For this particular case, $\sqrt{F_{\chi_{2}^2}^{-1}\left(1-\operatorname{P}_{\mathrm{FA}}\right)} = -2 \log \operatorname{P}_{\mathrm{FA}}$.
Moreover, it is well known that the Marcum Q function is strictly increasing in its first parameter, hence increasing \ac{SINR} $g_q$ contributes to an increase in $\operatorname{P}_{\mathrm{D}}^{(q)}$.
Therefore, we use the \ac{SINR} $g_q$ as a proxy due to its more tractable form, in order to influence an increase in the probability of detection.

%% file: actions/papr-explained.tex
Mathematically, the $\PAPR$ of a complex-valued signal vector over the $n^{th}$ transmit antenna samples can be defined as
\begin{equation}
	\label{eq:PAPR-equation}
\PAPR_n([\pmb{X}]_{n,:})
	=
	\frac{\max\nolimits_{\ell = 1 \ldots L}\vert [\pmb{X}]_{n,\ell} \vert^2}{1/L\sum\nolimits_{\ell = 1}^{L} \vert [\pmb{X}]_{n,\ell} \vert^2}, \forall n = 1\ldots N_T.
\end{equation}
Note that \eqref{eq:PAPR-equation} defines a per-antenna \ac{PAPR} constraint.
Under the assumption of identical transmit power of each antenna and that the total transmit power is normalized to $1$ (i.e. $\Vert \pmb{x} \Vert^2 = 1$, where $\pmb{x} = \ve(\pmb{X})$), we have that $\sum_{\ell = 1}^{L} \vert [\pmb{X}]_{n,\ell} \vert^2 = \frac{1}{N_T}$. Hence, constraining each $\PAPR_n([\pmb{X}]_{n,:})$ as 
\begin{equation}
\label{eq:PAPR-equation-2}
	\PAPR_n([\pmb{X}]_{n,:}) \leq \eta, \forall n = 1\ldots N_T,
\end{equation}
is equivalent to setting $\Vert \pmb{x} \Vert^2 = 1$ and $\max\nolimits_{\ell = 1 \ldots L}\vert [\pmb{X}]_{n,\ell} \vert^2 \leq \frac{\eta}{N_TL}$, or equivalently $\pmb{x}^H \pmb{F}_p \pmb{x} \leq \frac{\eta}{N_TL}$ for all $p = 1 \ldots N_TL$, where $\pmb{F}_p \in \mathbb{R}^{N_TL \times N_TL}$ is an all-zeros matrix, except for $1$ positioned at the $p^{th}$ diagonal component. 

%% file: actions/on-S-X.tex
Note that $\pmb{S}$ is the desired constellation that is intended to all users, i.e. $p^{th}$ row of $\pmb{S}$ is the constellation vector for the $p^{th}$ user. 
The precoding aspect is implicitly contained within $\boldsymbol{X}$. Indeed, an unconstrained minimization of \eqref{eq:EMUI} would give the zero-forcing solution $\pmb{H}^{\dagger} \pmb{S}$ for well-conditioned channels. However, with a constrained problem under sensing and \ac{PAPR}, this is not often the case.
In this sequel, the space-time sequence $\pmb{X}$ is synthesized directly at baseband and will be generated as solution given $\pmb{S},\pmb{H}$, using the formulation introduced in Section \ref{sec:opt-probs}.

%% file: actions/integrated-process.tex
\paragraph{Integrated process for prior information}
The proposed framework is designed for the tracking in order to have optimized target returns while maintaining communications. Within a \ac{CPI}, the \ac{DFRC} \ac{BS} adopts the following integrated process which can help answer a fundamental question, namely \emph{How can the \ac{BS} obtain prior information of the communication channel and target/interference information ?}
\begin{itemize}
	\item Initial phase: The \ac{DFRC} \ac{BS} transmits a sequence with no precoding or beamforming, and includes a preamble, in order to estimate the \acp{AoA} in the environment, where the users can use the preamble part to estimate their channel and feed it back to the \ac{DFRC} \ac{BS}. Methods for joint \ac{AoA}, doppler and delay estimation for the monostatic case can be reused from existing literature e.g. \cite{11275173} (c.f. Section II.A for $\beta_t = 0$ therein). Based on velocity estimates, the \ac{DFRC} \ac{BS} chooses to track targets above a certain velocity threshold, where a classification is done based on targets of interest, i.e. ($\hat{\theta}_1 \ldots \hat{\theta}_Q$) and interferers/clutter ($\hat{\bar{\theta}}_1 \ldots \hat{\bar{\theta}}_I$).
	Using the delay, Doppler and \ac{AoA} estimates, one can easily recover the \ac{RCS} \cite{11275173} via a simple inverse problem to estimate the two-way channel coefficients ($\hat{\gamma}_1 \ldots \hat{\gamma}_Q$) and ($\hat{\bar{\gamma}}_1 \ldots \hat{\bar{\gamma}}_I$). The communication users feed back channel estimates $\widehat{\pmb{H}}$. 
	
	\item Closed loop phase: Now, given the current  estimate of the communication channel and target/interference information, what is the optimal waveform to transmit next? The proposed DRIP framework uses the estimated values ($\hat{\theta}_1 \ldots \hat{\theta}_Q$), 
	($\hat{\bar{\theta}}_1, \ldots \hat{\bar{\theta}}_I$), ($\hat{\gamma}_1 \ldots \hat{\gamma}_Q$), ($\hat{\bar{\gamma}}_1 \ldots \hat{\bar{\gamma}}_I$) and $\hat{\pmb{H}}$ to generate the optimal waveform for the next transmission. As the radar \ac{SINR} is considered as metric for optimization, the receive radar \ac{SINR} is done on a target level not only for detection purposes (testing if the target still exists at angle $\hat{\theta}_q$ or not) but also to refine the angular estimation for the next transmission round and maintain communication with users.
\end{itemize}

%% file: sections/optimization-problems.tex
\subsection{\textcolor{black}{Connections with Information Theory}}
\label{sec:inf-theory}
\textcolor{black}{\input{actions/inf-theo}}

\subsection{\textcolor{black}{Problem Formulation}}
Now, we introduce \ac{DRIP} that takes into account multiple metrics related to sensing, communication and \ac{PAPR} restrictions.
Since one of our tasks is also to generate waveforms under a desired \ac{PAPR} value, the \ac{PAPR} on the waveform $\pmb{X}$ is restricted to be upper-bounded by a given level, say $\eta$. To this end, our \ac{DRIP} optimization problem can be formulated as follows
\begin{equation}
 \label{eq:problem1}
\textcolor{black}{\input{actions/DRIP-corrected}}
\end{equation}
\textcolor{black}{\input{actions/important1.tex}}
\textcolor{black}{\input{actions/epsilon-constrained}}The equivalent \ac{DRIP} optimization problem in consideration is 
\begin{equation}
 \label{eq:problem2}
\begin{aligned}
(\mathcal{P}_{\rm{DRIP}}):
\begin{cases}
\min\nolimits_{ \pmb{X}  , \pmb{U}}&  \Vert \pmb{H}\pmb{X} - \pmb{S} \Vert^2 \\
\textrm{s.t.}
 &  \Vert \pmb{x} \Vert^2 = 1, \quad  \pmb{x}^H \pmb{F}_p \pmb{x} \leq \frac{\eta}{N_TL} , \quad \forall p, \\
 & \pmb{x} \in \mathcal{B}_{\epsilon}(\pmb{x}_0),  g_q \geq \bar{g}_q, \quad \forall q,
\end{cases}
\end{aligned}
\end{equation}
\textcolor{black}{\input{actions/papr-explained-2}}
Before we proceed, we upper bound the \ac{MU} interference energy as follows
\begin{equation}
\label{eq:upper-bound-11}
	\begin{split}
		\Vert \pmb{H}\pmb{X} - \pmb{S} \Vert_F^2  &= \Vert \pmb{H} ( \pmb{X} - \pmb{H}^H(\pmb{H}\pmb{H}^H)^{-1}{\pmb{S}}) \Vert_F^2 \\
				 &\leq \Vert \pmb{H} \Vert_F^2 \Vert \pmb{x} - \pmb{x}_{\comm} \Vert^2,
	\end{split}
\end{equation}
where we invoked $\Vert \pmb{A}\pmb{B} \Vert_F^2 \leq\Vert \pmb{A} \Vert_F^2\Vert \pmb{B} \Vert_F^2 $ and $\pmb{x}_{\comm} = \ve(\pmb{H}^H(\pmb{H}\pmb{H}^H)^{-1}\pmb{S})$. 
 \begin{figure}[t]
	\centering
\includegraphics[width=0.75\linewidth]{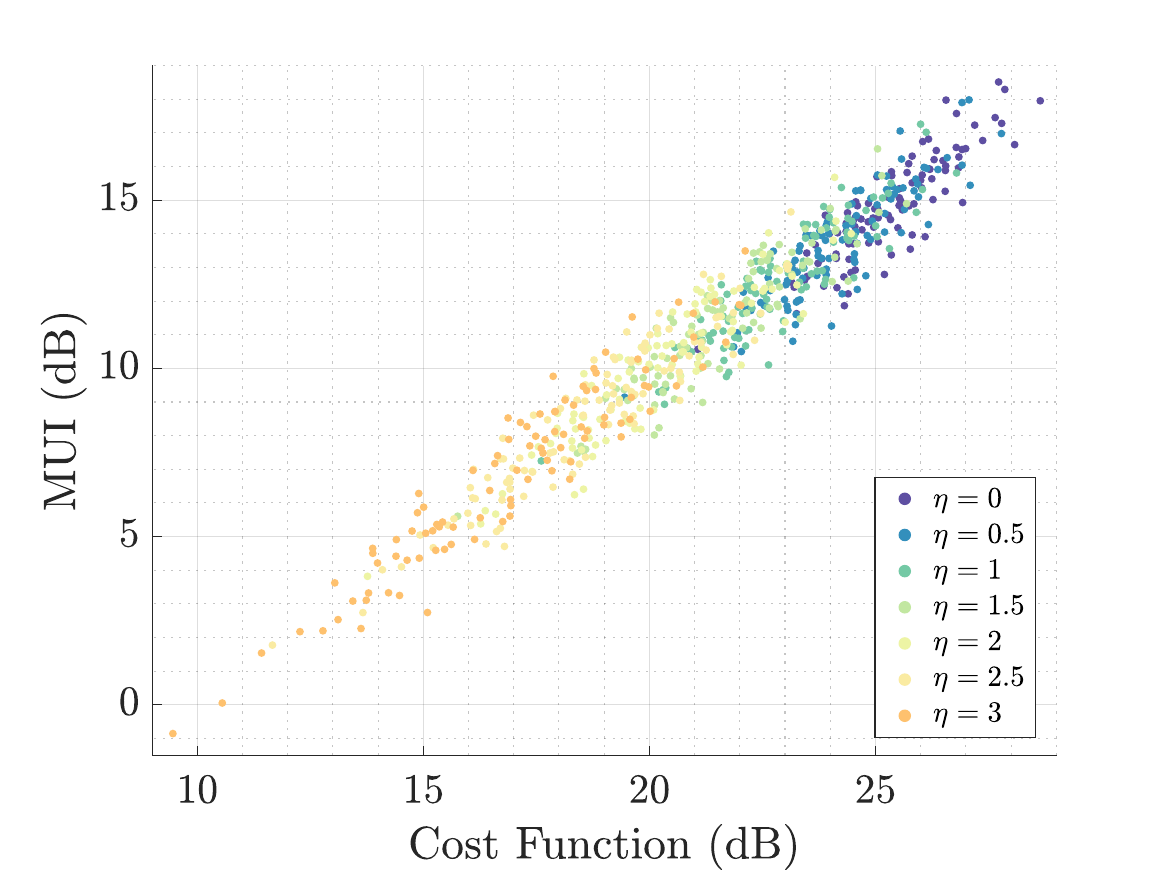}
 \caption{\textcolor{black}{The MUI behaviour of the actual cost function (upper bound), for $\epsilon = 2$, $N_T=12$, $N_R = 7$,  $L = 7$, $P=4$ and 16-QAM.}}
 \label{fig:MUI_vs_cost}
\end{figure}
\textcolor{black}{\input{actions/tighter-bound}}
Therefore, we replace the objective function in  \eqref{eq:problem2} with its upper bound in \eqref{eq:upper-bound-11}, resulting in a relaxed version of $(\mathcal{ P}_{\mathrm{DRIP}})$ as given below
\begin{equation}
 \label{eq:problem3}
\begin{aligned}
(\widetilde{\mathcal{P}}_{\rm{DRIP}}):
\begin{cases}
\min\nolimits_{ \pmb{x} ,  \pmb{U}}&  \Vert  \pmb{x} - \pmb{x}_{\comm} \Vert^2 \\
\textrm{s.t.}
 &  \Vert \pmb{x} \Vert^2 = 1, \quad  \pmb{x}^H \pmb{F}_p \pmb{x} \leq \frac{\eta}{N_TL} , \quad \forall p, \\
 & \pmb{x} \in \mathcal{B}_{\epsilon}(\pmb{x}_0),  g_q \geq \bar{g}_q, \quad \forall q.
\end{cases}
\end{aligned}
\end{equation}
It is worth expressing a fundamental difference between the introduced \ac{DRIP} design and our previously proposed \ac{SRIP} approach in \cite{10061453}.
\ac{SRIP} operates over the temporal dimension only, which is ideal for single-antenna \ac{ISAC} systems as the optimization is done solely over time domain.
In contrast, \ac{DRIP} is intended for a joint space-time \ac{ISAC} \textcolor{black}{sequences}, whereby space and time resources are jointly taken into account to yield an \ac{ISAC} \textcolor{black}{sequence} with a desired \ac{PAPR} level.
More specifically, the space-time \ac{ISAC} \textcolor{black}{sequence} generated by \ac{DRIP} not only preserves a similarity with respect to a radar waveform, while reducing \ac{MUI} for communication and maintaining a desirable \ac{PAPR} level, but is also capable of steering its energy towards multiple directions simultaneously to sense radar targets of interest, while rejecting interference and clutter along given directions, with the aid of receive beamforming.

\textcolor{black}{\input{actions/selectivity}}

\input{actions/versatile.tex}
Due to the non-convexity of $\widetilde{\mathcal{P}}_{\rm{DRIP}}$ in \eqref{eq:problem3}, in the next section, we design a suitable algorithm dedicated to generate \ac{DRIP}-type \textcolor{black}{sequences}.

%% file: actions/inf-theo.tex
From an information-theoretic perspective, the trade-off between the radar similarity parameter $\epsilon$ and the achievable communication sum-rate is analyzed through fundamental limits of a simplified version of the \ac{DRIP} framework by noting that the radar similarity constraint limits the maximum entropy of the source, thereby bounding the \ac{MI} available for communication.
In particular, the constraint $\mathcal{B}_{\epsilon}(\pmb{x}_0)$ restricts the energy of the deviation at the source to be at most $\epsilon$.
Let the transmitted sequence be decomposed as two components, i.e. deterministic and random
\begin{equation}
	\pmb{X} = \pmb{X}_0 + \pmb{W},
\end{equation}
where $\pmb{W}$ is the random data-carrying modulation. So $\mathcal{B}_{\epsilon}(\pmb{x}_0)$ also means $\|\pmb{W}\|_F^2 \le \epsilon$. Since $\pmb{X}_0$ is known, the \ac{MI} depends only on the random part $\pmb{W}$
\begin{equation*}
	I(\pmb{X}; \pmb{Y}_c) = I(\pmb{X}_0 + \pmb{W}; \pmb{H}(\pmb{X}_0 + \pmb{W}) + \pmb{Z}_c) = I(\pmb{W}; \pmb{H}\pmb{W} + \pmb{Z}_c),
\end{equation*}
which represents the achievable mutual information for a given error distribution.
The capacity is now
\begin{equation*}
	C(\epsilon) =  \max_{ \substack{ \pmb{R}_W \succeq \pmb{0} \\ \text{Tr}(\pmb{R}_W) \le \min(\epsilon,P_T - \|\pmb{X}_0\|_F^2) } }  \hspace{-0.8cm}L \log_2 \det \left( \pmb{I} + \frac{1}{\sigma_c^2} \pmb{H} \pmb{R}_W \pmb{H}^H \right),
\end{equation*}
where $\pmb{R}_W$ is the covariance of the random data-carrying part, and $P_T$ is the total power budget for $\pmb{X}$.
From here, one can see that $\epsilon$ plays the role of a power budget constraint on the data-carrying part of $\pmb{X}$ via $\pmb{W}$. Indeed, deriving a \ac{PAPR} constrained capacity in conjunction with the radar similarity could be done following Bussgang  approximations.

%% file: actions/DRIP-corrected.tex
\begin{aligned}
(\mathcal{P}_{\rm{DRIP}}):
\begin{cases}
\min\nolimits_{ \pmb{X} , \pmb{U}}&  \Vert \pmb{H}\pmb{X} - \pmb{S} \Vert^2 \\
\textrm{s.t.}
 &  \PAPR_n([\pmb{X}]_{n,:})\leq \eta, \forall n = 1 \ldots N_T,  \\ 
 & \pmb{x} \in \mathcal{B}_{\epsilon}(\pmb{x}_0) , g_q \geq \bar{g}_q, \quad \forall q,
\end{cases}
\end{aligned}

%% file: actions/important1.tex
where $g_q$ represents the radar \ac{SINR} for the $q$-th target (as defined in \eqref{eq:radar-sinr}) and $\bar{g}_q$ is the required \ac{SINR} threshold. We recall that $\PAPR_n$ is the \ac{PAPR} of the $n^{th}$ \ac{DRIP} sequence over antenna $n$, $\eta$ denotes the PAPR threshold, and $\pmb{x}_0$ is the vectorized version of the desired radar chirping signal.
Also, $\pmb{U} = \begin{bmatrix}
	\pmb{u}_{1,1} & \ldots & \pmb{u}_{1,L} & \ldots & 
	\pmb{u}_{Q,1} & \ldots & \pmb{u}_{Q,L}
\end{bmatrix}$ denotes the concatenation of all receive beamforming vectors, and $\pmb{x} = \ve(\pmb{X})$.

%% file: actions/epsilon-constrained.tex
A variant of $(\mathcal{P}_{\rm{DRIP}})$ can be formulated as a cost that weighs the \ac{MUI} and \ac{PAPR} metrics as
\begin{equation*}
	\begin{aligned}
(\mathcal{P}_{\rm{DRIP}}^w):
\begin{cases}
\min\limits_{ \pmb{X} , \pmb{U}}& w_1 \Vert \pmb{H}\pmb{X} - \pmb{S} \Vert^2 + w_2\max_n \PAPR_n([\pmb{X}]_{n,:}) \\
\textrm{s.t.}
 & \pmb{x} \in \mathcal{B}_{\epsilon}(\pmb{x}_0) ,  g_q \geq \bar{g}_q, \quad \forall q,
\end{cases}
\end{aligned}
\end{equation*}
for $w_1,w_2 \geq 0$.
which is referred to in multiobjective optimization literature as the \emph{weighting problem} \cite{miettinen1999nonlinear}, as opposed to the proposed problem $(\mathcal{P}_{\rm{DRIP}})$ in \eqref{eq:problem1} which is referred to as \emph{$\epsilon$-constrained} problem \cite{miettinen1999nonlinear}. As analyzed in \cite{miettinen1999nonlinear}, it is possible to find every Pareto optimal solution of any multiobjective optimization problem by \eqref{eq:problem1}, regardless of the convexity of the problem.
Interestingly, there are some intimate connections between both versions, i.e. between $(\mathcal{P}_{\rm{DRIP}}^w)$ and $(\mathcal{P}_{\rm{DRIP}})$, one of which is \cite{miettinen1999nonlinear} (c.f. Theorem 3.2.5) which tells us that:
Given a solution $\pmb{U}^*,\pmb{X}^*$ of weighting problem $(\mathcal{P}_{\rm{DRIP}}^w)$ and $w_1\geq 0,w_2\geq0$; if $w_{1}>0$ then $\pmb{U}^*,\pmb{X}^*$ is a solution of $(\mathcal{P}_{\rm{DRIP}})$ for $\eta$ set to $\max_n \PAPR_n([\pmb{X}^*]_{n,:})$.
This means that a solution of the jointly weighted cost function can be found using the proposed \ac{DRIP} formulation as in $(\mathcal{P}_{\rm{DRIP}})$ for a particular choice of $\eta$.

%% file: actions/papr-explained-2.tex
where as explain in Section \ref{subsec:metrics}, the per-antenna \ac{PAPR} constrains were replaced due to the identical per-antenna power constraints.

%% file: actions/tighter-bound.tex
The bound \eqref{eq:upper-bound-11} can be tightened as
\begin{equation}
\label{eq:tight-bound}
	\Vert \pmb{H}\pmb{X} - \pmb{S} \Vert_F^2
	\leq 
	\sigma_{\max }(\left.\pmb{H}\right|_{ \mathcal{R}(\pmb{X} - \pmb{X}_{\comm})})^2\|\pmb{X} - \pmb{X}_{\comm}\|^2,
\end{equation}
where for any two matrices $\pmb{A},\pmb{B}$, the $\sigma_{\max}(\pmb{A}|_{\mathcal{R}(\pmb{B})}) = \| \pmb{A} \pmb{P}_B \|_2$.
\emph{The bound in \eqref{eq:tight-bound} is tight and is achieved with equality when $\pmb{X} - \pmb{X}_{\comm}$ is rank-1.}
In Fig. \ref{fig:MUI_vs_cost}, we plot the actual \ac{MUI} $\Vert \pmb{H}\pmb{X} - \pmb{S} \Vert_F^2$ vs the cost function $\|\pmb{X} - \pmb{X}_{\comm}\|^2$ for sequences generated by \ac{DRIP}, which shows a clear trend that minimizing the adopted cost would also minimize on average the \ac{MUI}.

%% file: actions/selectivity.tex
At the transmitter side, the waveform $\mathbf{X}$ is designed to maximize the term $\sum\nolimits_{\ell = 1}^L \Vert  \pmb{u}_{q,\ell}^H \pmb{\Pi}(\theta_q) \pmb{X}_{:,\ell} \Vert^2$ in the numerator of the \ac{SINR} in equation \eqref{eq:radar-sinr}, which means the transmit signal itself is shaped to have high gain in the directions $\theta_q$. Also, $\mathbf{X}$ is constrained to minimize the energy reflected from known interfering directions $\bar{\theta}_{1} \ldots \bar{\theta}_{I}$, which can steer transmit nulls toward those directions.

At the receiver side, the optimization assigns a specific receive beamforming vector $\mathbf{u}_q$ for each direction $q$, which plays the role of a spatial filter. In particular, $\mathbf{u}_q$ combines the signals from all $N_R$ receive antennas with specific weights in order to look in direction $\theta_q$, while attempting to minimize energy in the directions of interferers, $\bar{\theta}_{1} \ldots \bar{\theta}_{I}$. 
The level $\bar{g}_q$ also influences the level of selectivity towards direction $\theta_q$.

%% file: actions/versatile.tex
\textcolor{black}{\textbf{Discussion:}}
\textcolor{black}{We coin the term \emph{"versatile"} to emphasize that the \ac{DRIP} sequences are able to adapt to different \ac{ISAC} and hardware functionalities given the input $(\epsilon,\eta, \lbrace \bar{g}_q \rbrace,\pmb{H},\pmb{S})$.}
\textcolor{black}{The $\epsilon$ acts as a parameter (for time-domain) that moves the system between sensing and communication regions. In particular, when $\epsilon$ is small, the waveform is forced to closely resemble a radar chirp, which then maximizes sensing performance. As $\epsilon$ loosens, the waveform would entail more degrees-of-freedom to prioritize \ac{MUI} minimization for communications.}
\textcolor{black}{The parameter $\bar{g}_q$ acts as a parameter (for space-domain) that allows the sequence to focus on certain direction $\theta_q$ while rejecting power in directions $\bar{\theta}_1 \ldots \bar{\theta}_I$. Also, as we related in Section \ref{sec:system-model}, paragraph (b), the increase in $g_q$ (through its lower bound $\bar{g}_q$) also increase the probability of detection of target $q$}
\textcolor{black}{In terms of \ac{PAPR}, the \ac{DRIP} sequence can be generated to respect a given \ac{PAPR} level $\eta$. In particular, in scenarios where non-linear amplifiers require low backoff to maximize efficiency, $\eta$ can be set close to $0 \dB$. Once $\eta$ is relaxed, one can expect lower \ac{MUI} values which then allows higher data rates.}
\textcolor{black}{In short, the proposed framework yields a versatile continuum of solutions, where for any given set of parameters defining the constraints, namely $(\epsilon,\eta, \lbrace \bar{g}_q \rbrace,\pmb{H},\pmb{S})$, $\left(\tilde{\mathcal{P}}_{\mathrm{DRIP}}\right)$ generates a corresponding DRIP sequence $\pmb{X}$ and its corresponding receive beamformers $\pmb{U}$. In other words, given $(\epsilon, \eta, \{\Bar{g}_q\}, \boldsymbol{H}, \boldsymbol{S})$ one DRIP sequence $\pmb{X}$ and one set of beamformers $\pmb{U}$ is generated.}

%% file: sections/optimization01.tex
The first block of the \ac{BCCD} method optimizes with respect to $\pmb{U}$.
Before we proceed, note that the \ac{SINR} in \eqref{eq:radar-sinr} can be written in the form of a generalized Rayleigh quotient term as 
\vspace{-0.2cm}
\textcolor{black}{\input{actions/gq-corrected}}
where 
\begin{align}
	\pmb{T}_{1,q}
	&=
	\sigma_q^2
	\big( \pmb{I}_L \otimes \pmb{\Pi}(\theta_q) \big) \pmb{x}
		\pmb{x}^H
	\big( \pmb{I}_L \otimes \pmb{\Pi}(\theta_q) \big)^H,\\
\begin{split}
	\pmb{T}_{2,q}
	&=
	\sum\nolimits_{q' \neq q}
	\sigma_{q'}^2
	\big( \pmb{I}_L \otimes \pmb{\Pi}(\theta_{q'}) \big) \pmb{x}
		\pmb{x}^H
	\big( \pmb{I}_L \otimes \pmb{\Pi}(\theta_{q'}) \big)^H	 \\
	&+
	\sum\nolimits_{i}
	\bar{\sigma}_{i}^2
	\big( \pmb{I}_L \otimes \pmb{\Pi}(\bar{\theta}_{i}) \big) \pmb{x}
		\pmb{x}^H
	\big( \pmb{I}_L \otimes \pmb{\Pi}(\bar{\theta}_{i}) \big)^H
	+
	\sigma_r^2 \pmb{I} \label{eq:T2q},
\end{split}
\end{align}
and $\pmb{u}_q = \begin{bmatrix}
	\pmb{u}_{q,1}^T & \ldots & \pmb{u}_{q,L}^T\end{bmatrix}^T$.
Next, observe that the only quantity depending on $\pmb{u}_q$ is $g_q$. So, adopting an alternating optimization approach, fixing $\pmb{x}$, the optimization problem in \eqref{eq:problem3} should aim at maximizing $g_q$. To this extent, we have the following $Q$ independent optimization problems to be solved
\begin{equation}
\label{eq:uq-maximization}
\widehat{\pmb{u}}_q
=
	\argmax_{\pmb{u}_q}
	\frac{\pmb{u}_q^H \pmb{T}_{1,q} \pmb{u}_q}{\pmb{u}_q^H \pmb{T}_{2,q} \pmb{u}_q},
	\quad 
	\forall q = 1 \ldots Q.
\end{equation}
Due to the fractional nature of the maximization in \eqref{eq:uq-maximization}, a possible workaround is by holding the numerator constant, and minimizing the denominator, namely
\begin{equation}
 \label{eq:problemu}
\begin{aligned}
(\widetilde{\mathcal{P}}^u_{\rm{DRIP}}):
\begin{cases}
\min\nolimits_{\pmb{u}_q}&  \pmb{u}_q^H \pmb{T}_{2,q} \pmb{u}_q \\
\textrm{s.t.}
 &  \pmb{u}_q^H\big( \pmb{I}_L \otimes \pmb{\Pi}(\theta_{q}) \big) \pmb{x} = \zeta, \\ 
\end{cases}
\end{aligned}
\end{equation}
where $\zeta$ is a constant. Using optimization theory, the solution of $(\widetilde{\mathcal{P}}^u_{\rm{DRIP}})$ in \eqref{eq:problemu} above is given as 
\begin{equation}
	\label{eq:BFer-update}
	\widehat{\pmb{u}}_q
	=
	\frac{\pmb{T}_{2,q}^{-1}( \pmb{I}_L \otimes \pmb{\Pi}(\theta_{q}) \big) \pmb{x}}{\pmb{x}^H( \pmb{I}_L \otimes \pmb{\Pi}(\theta_{q}) \big)^H\pmb{T}_{2,q}^{-1}( \pmb{I}_L \otimes \pmb{\Pi}(\theta_{q}) \big) \pmb{x}}
\end{equation}
After obtaining $\widehat{\pmb{u}}_q$, the radar \ac{SINR} is evaluated at this point, and hence the problem in \eqref{eq:problem3} becomes  
\begin{equation}
 \label{eq:problem3-x}
\begin{aligned}
(\widetilde{\mathcal{P}}^x_{\rm{DRIP}}):
\begin{cases}
\min\nolimits_{ \pmb{x} }&  \Vert  \pmb{x} - \pmb{x}_{\comm} \Vert^2 \\
\textrm{s.t.}
 &  \textcolor{black}{\Vert \pmb{x} \Vert^2 = 1,} \quad  \pmb{x}^H \pmb{F}_p \pmb{x} \leq \frac{\eta}{N_TL} , \quad \forall p \\
 & \pmb{x} \in \mathcal{B}_{\epsilon}(\pmb{x}_0), \quad  \widehat{g}_q \geq \bar{g}_q, \quad \forall q,
\end{cases}
\end{aligned}
\end{equation} 
where $\widehat{g}_q = \frac{\widehat{\pmb{u}}_q^H \pmb{T}_{1,q} \widehat{\pmb{u}}_q}{\widehat{\pmb{u}}_q^H \pmb{T}_{2,q} \widehat{\pmb{u}}_q}$, $\forall q = 1 \ldots Q$. 
\textcolor{black}{\input{actions/why-bccd}} Now, we introduce the following property\\
\textbf{Property 1}: Problem $(\widetilde{\mathcal{P}}^x_{\rm{DRIP}})$ in  \eqref{eq:problem3-x} is a \ac{QCQP} optimization problem with $N_TL + Q + 3$ inequality quadratic constraints and no equality constraints ,i.e.
\begin{equation}
 \label{eq:problem-qcqp}
\begin{aligned}
(\mathcal{S}_{\rm{QCQP}}):
\begin{cases}
\min\nolimits_{ \pmb{x}_r }&  \pmb{x}_r^T \pmb{P}_0 \pmb{x}_r +  \pmb{q}_0^T \pmb{x}_r \\
\textrm{s.t.}
 &  \pmb{x}_r^T \pmb{P}_i \pmb{x}_r +  \pmb{q}_i^T \pmb{x}_r  \leq  r_i , \quad \forall i = 1 \ldots m.
\end{cases}
\end{aligned}
\end{equation}
\textit{where $m=N_TL + Q + 3$. 
The optimization variable is real-valued $\pmb{x}_r \in \mathbb{R}^{2N_T L \times 1}$.
In addition, $\pmb{P}_i \in \mathbb{R}^{2N_T L \times 2N_T L }$,$\pmb{q}_i \in \mathbb{R}^{2N_T L \times 1}$, and  $ r_i \in \mathbb{R}$ for all $i $. }\\
\textbf{Proof}: See Appendix \ref{appendix:proof-QCQP}.  \\
Next, we apply non-negative slack variables onto the inequality constraints so that the inequalities are expressed as linear instead of quadratic. So, problem $(\mathcal{S}_{\rm{QCQP}})$ can be equivalently represented by the following 
\begin{equation}
 \label{eq:problem-slacked}
\begin{aligned}
(\mathcal{S}_{\rm{QCQP}}'):
\begin{cases}
\min\nolimits_{ \pmb{x}_r }&  \pmb{x}_r^T \pmb{P}_0 \pmb{x}_r +  \pmb{q}_0^T \pmb{x}_r \\
\textrm{s.t.}
 &  \pmb{x}_r^T \pmb{P}_i \pmb{x}_r +  \pmb{q}_i^T \pmb{x}_r  + \varphi_i =  r_i , \ \forall i = 1 \ldots m,\\\
 & 	\varphi_i \geq 0, \quad \forall i = 1 \ldots m. 
\end{cases}
\end{aligned}
\end{equation}
Stacking the slack variables in a vector, say $\pmb{\varphi}
	=
	\begin{bmatrix}
		\varphi_1 & \ldots & \varphi_m
	\end{bmatrix}^T$, 
the augmented Lagrangian associated with the problem in \eqref{eq:problem-slacked} is given as 
\begin{equation}
\label{eq:l_rho_expression}
\begin{split}
	\mathcal{L}_\rho(\pmb{x}_r, \pmb{\varphi},\pmb{\lambda})
	&=
	\pmb{x}_r^T \pmb{P}_0 \pmb{x}_r +  \pmb{q}_0^T \pmb{x}_r
	\\ &+
	\sum\nolimits_i
	\lambda_i
	(\pmb{x}_r^T \pmb{P}_i \pmb{x}_r +  \pmb{q}_i^T \pmb{x}_r  + \varphi_i -  r_i)  
	\\ &+ 
	\frac{\rho}{2}
	\sum\nolimits_i
	(\pmb{x}_r^T \pmb{P}_i \pmb{x}_r +  \pmb{q}_i^T \pmb{x}_r  + \varphi_i -  r_i)^2,
\end{split}
\end{equation}
where $\pmb{\varphi}$ and $\pmb{\lambda}
	=
	\begin{bmatrix}
		\lambda_1 & \ldots & \lambda_m
	\end{bmatrix}^T$ are dual variables.
In addition, $\rho \geq 0$ is a penalty parameter 
We first optimize with respect to $\pmb{\varphi}$, i.e. we solve the following
\begin{equation}
	\widehat{\pmb{\varphi}}
	=
	\argmin\nolimits_{\pmb{\varphi}}
	\mathcal{L}_\rho(\pmb{x}_r, \pmb{\varphi},\pmb{\lambda}),
\end{equation}
which is solved by taking the gradient of $\mathcal{L}_\rho(\pmb{x}_r, \pmb{\varphi},\pmb{\lambda})$ with respect to $\varphi_i$, namely
\begin{equation}
	\frac{\partial}{\partial \varphi_i}
	\mathcal{L}_\rho(\pmb{x}_r, \pmb{\varphi},\pmb{\lambda})
	=
	\rho
	(\pmb{x}_r^T \pmb{P}_i \pmb{x}_r +  \pmb{q}_i^T \pmb{x}_r  + \varphi_i -  r_i)
	+
	\lambda_i,
\end{equation}
if $\varphi_i \geq 0$, and zero otherwise.
Setting the gradient to zero, we get the following update equation
\begin{equation}
	\label{eq:varphi_i-expression}
\begin{split}
	 \varphi_i^{(n)}&=\Big[ -\frac{\lambda_i}{\rho}
- \pmb{x}_r^T \pmb{P}_i \pmb{x}_r 
-  \pmb{q}_i^T \pmb{x}_r
	+r_i \Big]^+.
\end{split}
\end{equation}
\vspace{-0.2cm}
Replacing \eqref{eq:varphi_i-expression} in \eqref{eq:l_rho_expression}, we get
\begin{equation}
\label{eq:L-rho-hat}
\begin{split}
	\widehat{\mathcal{L}}_\rho(\pmb{x}_r,\pmb{\lambda})
	&=
	\pmb{x}_r^T \pmb{P}_0 \pmb{x}_r 
	+  \pmb{q}_0^T \pmb{x}_r - \sum\nolimits_i 
	 \frac{\lambda_i^2}{2\rho}
	\\ &+
	\frac{\rho}{2}
	\sum\nolimits_i 
	\left( \left[\frac{\lambda_i}{\rho}
	+\pmb{x}_r^T \pmb{P}_i \pmb{x}_r
	+ \pmb{q}_i^T \pmb{x}_r
	- r_i\right]^+ \right)^2.
\end{split}
\end{equation}
The optimization with respect to $\pmb{x}_r$ is achieved by solving 
\begin{equation}
\label{eq:minimize-L-rho}
	\pmb{x}_r^{(n)}=\arg \min _{\pmb{x}_r} \widehat{\mathcal{L}}_\rho(\pmb{x}_r,\pmb{\lambda}^{(n-1)}).
\end{equation}
The optimization problem in \eqref{eq:minimize-L-rho} is an unconstrained nonlinear optimization problem.
As the Hessian is too expensive to compute at every iteration, we resort to a Quasi-Newton type method which can approximate the Hessian matrix via successive iterates based on the gradients of $\widehat{\mathcal{L}}_\rho(\pmb{x}_r,\pmb{\lambda}^{(n-1)})$ with respect to $\pmb{x}_r$.
For this, we propose to solve \eqref{eq:minimize-L-rho} using the \ac{BFGS} algorithm.
Finally, the dual variables are updated as
\begin{equation*} 
\lambda_i^{(n)}  =\lambda_i^{(n-1)} + \rho
\left( (\pmb{x}_r^{(n)})^T \pmb{P}_i \pmb{x}_r^{(n)} +  \pmb{q}_i^T \pmb{x}_r^{(n)}  + \varphi_i^{(n)} -  r_i
\right),
\end{equation*}
which can be re-expressed as 
\begin{equation} 
\label{eq:lambda_i_n}
\lambda_i^{(n)} =\left[ \lambda_i^{(n-1)}+\rho\left((\pmb{x}_r^{(n)})^T \pmb{P}_i \pmb{x}_r^{(n)} +  \pmb{q}_i^T \pmb{x}_r^{(n)}-r_i\right)\right]^+ .
\end{equation}
The steps are given in \textbf{Algorithm \ref{alg:alg1}}.
\textcolor{black}{\input{actions/on-varphi}}

%% file: actions/gq-corrected.tex
\begin{equation}
	g_q
	=
	\frac{\pmb{u}_q^H \pmb{T}_{1,q} \pmb{u}_q}{\pmb{u}_q^H \pmb{T}_{2,q}\pmb{u}_q},
\end{equation}

%% file: actions/why-bccd.tex
The \ac{BCCD} framework is selected to decompose the non-convex optimization problem, $(\widetilde{\mathcal{P}}_{\text {DRIP}})$, into two analytically tractable sub-problems: $(\widetilde{\mathcal{P}}_{\mathrm{DRIP}}^u)$ for the receive beamformers and $(\widetilde{\mathcal{P}}_{\text {DRIP }}^x)$ for the transmit waveform. In essence, the approach directly addresses the computational intractability arising from the complex coupling between the beamformers $\mathbf{U}$ and the waveform $\mathbf{x}$ within the SINR constraints. Moreover, the algorithm decouples the variables, fixing one to optimize the other in an iterative fashion, allowing us to transform the original joint optimization $(\widetilde{\mathcal{P}}_{\text {DRIP}})$ into a sequence of relatively more tractable steps.

%% file: actions/on-varphi.tex
Note that the updates of $\pmb{\varphi}$ in  \eqref{eq:varphi_i-expression} have been injected in the original augmented Lagrangian expression in \eqref{eq:l_rho_expression} which then leads to the minimization criterion in \eqref{eq:l_rho_expression} that is optimized using \ac{BFGS} as described in the inner-loop steps of \textbf{Algorithm 1}.

%% file: sections/algo-description-1.tex
\begin{algorithm}[H]
\footnotesize
\caption{\ac{BCCD} algorithm tailored for \ac{DRIP}}\label{alg:alg1}
\begin{algorithmic}[1]
\STATE {\textsc{input}: $\pmb{x}_{\comm}, \pmb{x}_{0},\eta,\epsilon,\rho$}
\STATE {\textsc{set}:} $\pmb{P}_0 = \pmb{P}_1= \pmb{I}$, $\pmb{P}_2 = - \pmb{I}$, $\pmb{P}_{3} = \pmb{I}$, \\
 \hspace{0.5cm} $\pmb{q}_{0} = -2\phi(\pmb{x}_{\comm})$, $\pmb{q}_1 = \pmb{q}_2 = \pmb{0}$, $\pmb{q}_{3} = -2\phi(\pmb{x}_0)$\\
 \hspace{0.5cm} $r_1 = 1$ and $r_2 = -1$, $r_{3} =  \Vert \pmb{x}_0 \Vert^2 - \epsilon^2$.\\
\hspace{0.5cm} {\textsc{for}} $p = 1 \ldots N_TL$ \\
\hspace{1cm} $\pmb{P}_{p+3} = \phi(\pmb{F}_p)$, $\pmb{q}_{p+3} = \pmb{0}$, $r_{p+3} = \frac{\eta}{N_TL}$.\\
\STATE {\textsc{initialize}:} 
\STATE \hspace{0.5cm} $\pmb{x}^{(0)} = \pmb{x}_0$, $\pmb{x}_r^{(0)} = \phi(\pmb{x}^{(0)})$.
\STATE {\textsc{for}} $k =1 \ldots K_{\mathrm{iter}}$ {\tt{(outer-loop)}}
\STATE \hspace{0.5cm} {\textsc{for}} $q = 1 \ldots Q$\\
\hspace{1cm} Following \eqref{eq:BFer-update}, update the \textcolor{black}{beamformers} as 
\begin{equation*}
	\widehat{\pmb{u}}_q^{(k)}
	=
	\frac{\pmb{T}_{2,q}^{-1}( \pmb{I}_L \otimes \pmb{\Pi}(\theta_{q'}) \big) \pmb{x}^{(k-1)}}{{\pmb{x}^{(k-1)}}^H( \pmb{I}_L \otimes \pmb{\Pi}(\theta_{q'})^H\pmb{T}_{2,q}^{-1}( \pmb{I}_L \otimes \pmb{\Pi}(\theta_{q'}) \big) \pmb{x}^{(k-1)}}
\end{equation*}
\\\hspace{1cm} Using $\widehat{\pmb{u}}_q^{(k)}$, and following \eqref{eq:S1q} and \eqref{eq:S2q}, compute
\begin{align*}
	\hspace{0.5cm}\pmb{S}_{1,q}^{(k)} 
	&=
	\sigma_q^2
	\big( \pmb{I}_L \otimes \pmb{\Pi}(\theta_q) \big)^H
	\widehat{\pmb{u}}_q^{(k)}
	(\widehat{\pmb{u}}_q^{(k)})^H
	\big( \pmb{I}_L \otimes \pmb{\Pi}(\theta_q) \big) \\
	\begin{split}
	\pmb{S}_{2,q}^{(k)} 
	&=
	\sum\nolimits_{q' \neq q}
	\sigma_{q'}^2
	\big( \pmb{I}_L \otimes \pmb{\Pi}(\theta_{q'}) \big)^H
	\widehat{\pmb{u}}_q^{(k)}
	(\widehat{\pmb{u}}_q^{(k)})^H
	\big( \pmb{I}_L \otimes \pmb{\Pi}(\theta_{q'}) \big) 
			 \\
	&+
	\sum\nolimits_{i}
	\bar{\sigma}_{i}^2
	\big( \pmb{I}_L \otimes \pmb{\Pi}(\bar{\theta}_{i}) \big)^H
	\widehat{\pmb{u}}_q^{(k)}
	(\widehat{\pmb{u}}_q^{(k)})^H
	\big( \pmb{I}_L \otimes \pmb{\Pi}(\bar{\theta}_{i}) \big) 
	\end{split}
\end{align*}
\\\hspace{1cm} Fix $\pmb{q}_{N_TL + 3 +  q} = \pmb{0}$. Given $\pmb{S}_{1,q}^{(k)}, \pmb{S}_{2,q}^{(k)},\widehat{\pmb{u}}_q^{(k)}$, compute 
\begin{align*}
	\pmb{P}_{N_TL + 3 +q} &=\phi(\bar{g}_q \pmb{S}_{2,q}^{(k)} - \pmb{S}_{1,q}^{(k)}), \\
	r_{N_TL + 3 +  q} &=-  \bar{g}_q  \Vert  \widehat{\pmb{u}}_q^{(k)}  \Vert^2
\end{align*}
\vspace{-0.5cm}
\STATE \hspace{0.5cm} {\textsc{end for}}
\STATE \hspace{0.5cm} Initialize $\pmb{\lambda}^{(0)} = \pmb{0}$ 
\STATE \hspace{0.5cm} {\textsc{for}} $n = 1 \ldots N_{\mathrm{iter}}$ {\tt{(inner-loop)}}\\
\hspace{1cm} Obtain $\pmb{x}_r^{(n)}$ by solving \eqref{eq:minimize-L-rho} by the \ac{BFGS} algorithm. \\
\hspace{1cm} Update $\lambda_i^{(n)}$ via \eqref{eq:lambda_i_n}.
\STATE \hspace{0.5cm} {\textsc{end for}}
\STATE \hspace{0.5cm} Set $\pmb{x}^{(k)} = \phi^{-1}(\pmb{x}_r^{(N_{\mathrm{iter}})})$.
\STATE \textbf{return}  $\pmb{x}^{(k)}$.
\STATE {\textsc{end for}}
\end{algorithmic}
\end{algorithm}

%% file: sections/convergence-analysis.tex
In this section, we perform a convergence analysis of the proposed \ac{BCCD} algorithm dedicated to generate \ac{DRIP}-type waveforms.
We start by proving convergence of the inner-loop, which is detailed in the following theorem.

\begin{theorem}
\label{theorem:theorem1} (\textit{Inner-loop convergence})
If for every $\epsilon_1 > 0$ and $\epsilon_2 > 0$ we can find an iteration $N$, such that for all iterations $n > N$, we can bound $\left| (\pmb{x}_r^{(n+1)})^T \pmb{P}_i \pmb{x}_r^{(n+1)} - \bar{\pmb{x}}_r^T \pmb{P}_i \bar{\pmb{x}}_r \right| \leq \epsilon_1 $ 
and 
$\left| \pmb{q}_i^T \pmb{x}_r^{(n+1)}- \pmb{q}_i^T \bar{\pmb{x}}_r \right| \leq \epsilon_2,$ $\forall i$, then the algorithm is guaranteed convergence towards $\bar{\pmb{x}}_r$.
\end{theorem}

\textbf{Proof}: See \textbf{Appendix \ref{appendix:station-QCQP}}.

Theorem \ref{theorem:theorem1} only proves convergence of the inner-loop of \textbf{Algorithm \ref{alg:alg1}}, which is necessary but not a sufficient condition for the convergence of the algorithm.
To prove convergence of the algorithm, we must prove convergence of the outer-loop, which is given in the next theorem.

\begin{theorem}
\label{theorem:theorem2} (\textit{Outer-loop convergence})
The \ac{BCCD} iterative method designed in \textbf{Algorithm \ref{alg:alg1}}, is guaranteed to converge towards a feasible \ac{DRIP} solution, i.e. the converged waveform should satisfy the constraints of optimization problem $(\widetilde{\mathcal{P}}_{\rm{DRIP}})$ given in  \eqref{eq:problem3}.
\end{theorem}

\textbf{Proof:} See \textbf{Appendix \ref{appendix:outer-convergence}}.

\textcolor{black}{\input{actions/convergence-rate}}

%% file: actions/convergence-rate.tex
So far, the analysis above proves convergence towards a \ac{DRIP} solution. The following presents a worst-case convergence rate of the proposed \ac{BCCD} for \ac{DRIP}:

\textbf{Property 1}: (\textit{On convergence rate \& conditions}) Since the problem $(\widetilde{\mathcal{P}}_{\rm{DRIP}})$ in \eqref{eq:problem3} is built from finite unions and intersections of polynomial equalities and inequalities, it is formally classified as a \textit{semi-algebraic function} (see \cite{attouch2010proximal} for a formal definition of semi-algebraic functions and semi-algebraic sets).
It immediately follows that there exists some $\theta$ for which problem \eqref{eq:problem3} satisfies the Kurdyka-Łojasiewicz property with $\varphi(s)=c s^{1-\theta}$, for some $\theta \in[0,1)$ and some $c>0$\cite{attouch2010proximal}. Following \cite{attouch2010proximal} (c.f. Theorem 3.4), this means the rate of convergence depends on the value of $\theta$. As deriving the exact value of $\theta$ can be intractable, the Theorem reveals that the worst-case convergence rate is sublinear because when $\theta \in\left(\frac{1}{2}, 1\right)$ (worst case), then there exists $c>0$, such that
\begin{equation}
	\left\|\left(\pmb{x}^{(k)} , \pmb{U}^{(k)}\right)-\left(\pmb{x}^{(\infty)} , \pmb{U}^{(\infty)}\right)\right\| \leq c k^{-(1-\theta) /(2 \theta-1)}.
\end{equation}

%% file: actions/complexity-analysis.tex
In this section, we analyze the per-iteration complexity of \textbf{Algorithm \ref{alg:alg1}}. 
Forming $\boldsymbol{T}_{2, q}^{(k)}$ using \eqref{eq:T2q} costs $\mathcal{O}\left( (Q+I)(N_T N_R^2 L^3 + N_R N_T^2 L^3)  \right)$ and hence its inverse,  $[\boldsymbol{T}_{2, q}^{(k)}]^{-1}$,  requires $\mathcal{O}\left( (Q+I)^3(N_T N_R^2 L^3 + N_R N_T^2 L^3)^3  \right)$ operations. Moroever, implementing $Q$ inverses independently costs $\mathcal{O}\left( Q(Q+I)^3(N_T N_R^2 L^3 + N_R N_T^2 L^3)^3  \right)$.
 Similarly, forming $\boldsymbol{S}_{1, q}^{(k)}$ costs $\mathcal{O}(N_T N_R^2 L^3 + N_R N_T^2 L^3 )$, hence the overall cost to implement all $\boldsymbol{S}_{1, q}^{(k)}$'s over $q$ is $\mathcal{O}\left(Q(N_T N_R^2 L^3 + N_R N_T^2 L^3) \right)$.
%
%
%
%
%
Furthermore, since $\boldsymbol{S}_{2, q}^{(k)}$ follows a similar form of $S_{1,q}$, one would only need to recompute the interference part (second summand appearing in \eqref{eq:S2q}) because
\begin{equation}
	\pmb{S}_{2,q} 
	=
	\sum\nolimits_{q' \neq q}
	\pmb{S}_{1,q}+
	\sum\nolimits_{i}
	\bar{\sigma}_{i}^2
	\big( \pmb{I}_L \otimes \pmb{\Pi}(\bar{\theta}_{i}) \big)^H
	\widehat{\pmb{u}}_q
	\widehat{\pmb{u}}_q^H
	\big( \pmb{I}_L \otimes \pmb{\Pi}(\bar{\theta}_{i}) \big).
\end{equation}
The cost of $\pmb{S}_{2,q}$ is hence dominated by the second term, viz. $\mathcal{O}\left(I(N_T N_R^2 L^3 + N_R N_T^2 L^3) \right)$, and performing the computation over all $q$ gives a total cost of $\mathcal{O}\left(IQ(N_T N_R^2 L^3 + N_R N_T^2 L^3) \right)$.
Forming $\boldsymbol{P}_{N_T L+3+q}$ costs $\mathcal{O}(N_TL)$ and hence the total cost over all $\boldsymbol{P}_{N_T L+3+q}$ is $\mathcal{O}(QN_TL)$. Also, computing $r_{N_T} L+3+q$ costs of simply $\mathcal{O}(N_RL)$ to compute the norm and so the overall cost for all $r_{N_T} L+3+q$ is $\mathcal{O}(QN_RL)$.
In addition, solving for $\boldsymbol{x}_r^{(n)}$ using \ac{BFGS} costs $\mathcal{O}(N_T^2 L^2)$(see \cite{nocedal2006numerical} c.f. Chapter 6), hence iterating $N_{\mathrm{iter}}$ times costs $\mathcal{O}\left( N_{\mathrm{iter}} N_T^2 L^2 \right)$. 
%
%
%
Also, updating $\lambda_i^{(n)}$ is dominated by $\mathcal{O}(N_T^2 L^2)$ and hence iterating $N_{\mathrm{iter}}$ times leads to a total cost of $\mathcal{O}\left( N_{\mathrm{iter}} N_T^2 L^2 \right)$.
Finally, summing over all the previous operations gives an asymptotic computational complexity as
\begin{equation}
	T_{\mathrm{DRIP}}
	=
	\mathcal{O}\left( Q(Q+I)^3(N_T N_R^2 L^3 + N_R N_T^2 L^3)^3  \right),
\end{equation} 
which is dominated by the inverses of $\pmb{T}_{2,q}$. 

%% file: sections/simulation-results.tex
In this section, numerical simulation results for characterizing the performance and trade-offs of the proposed DFRC-waveform design are presented. 
The number of receive antennas is fixed to $N_R = 7$ antennas.
The number of radar targets is $Q = 2$ targets and the number of interferers is set to $I = 1$.
Furthermore, the communication channel $\pmb{H}$ is Rayleigh \ac{i.i.d} distributed. 
The radar noise variance is set to $\sigma_v^2 = 0.01$.
Moreover, The \ac{BFGS} number of iterations are set to $50$. 
The penalty factor is set to $\rho = 10$.
The utilized chirp for sensing is a \ac{LFM} chirp.

\begin{figure}[t]
 \centering
 \includegraphics[width=0.75\linewidth]{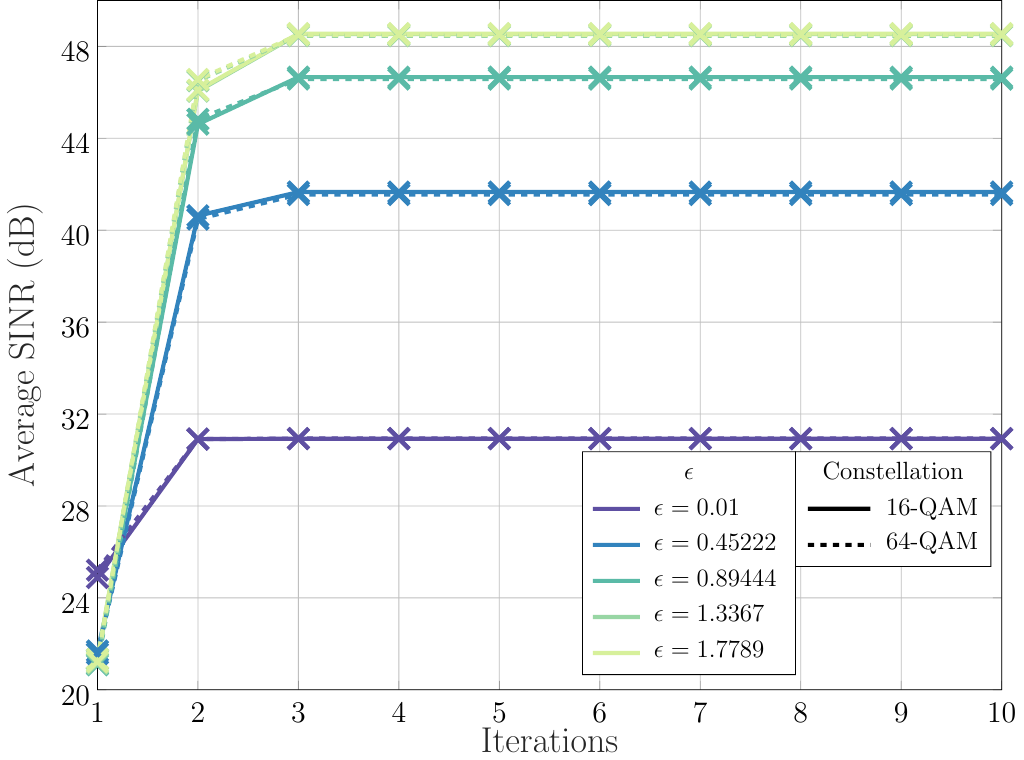}
 \caption{Average radar \ac{SINR} of the first target  over iterations, when two targets are presents at $\theta_1 = 10^\circ$ and $\theta_2 = 30^\circ$. The number of Tx antennas is $N_T= 12$, the signal length $L = 7$, the number of communication users $P = 4$ users, and $\eta = 2.5 \dB$. The radar \ac{SINR} constraint levels are set to $\bar{g}_1 = \bar{g}_2 = 20\dB$.}
 \label{fig:SINR-iter}
\end{figure}

\begin{figure}[t]
 \centering
 \includegraphics[width=0.75\linewidth]{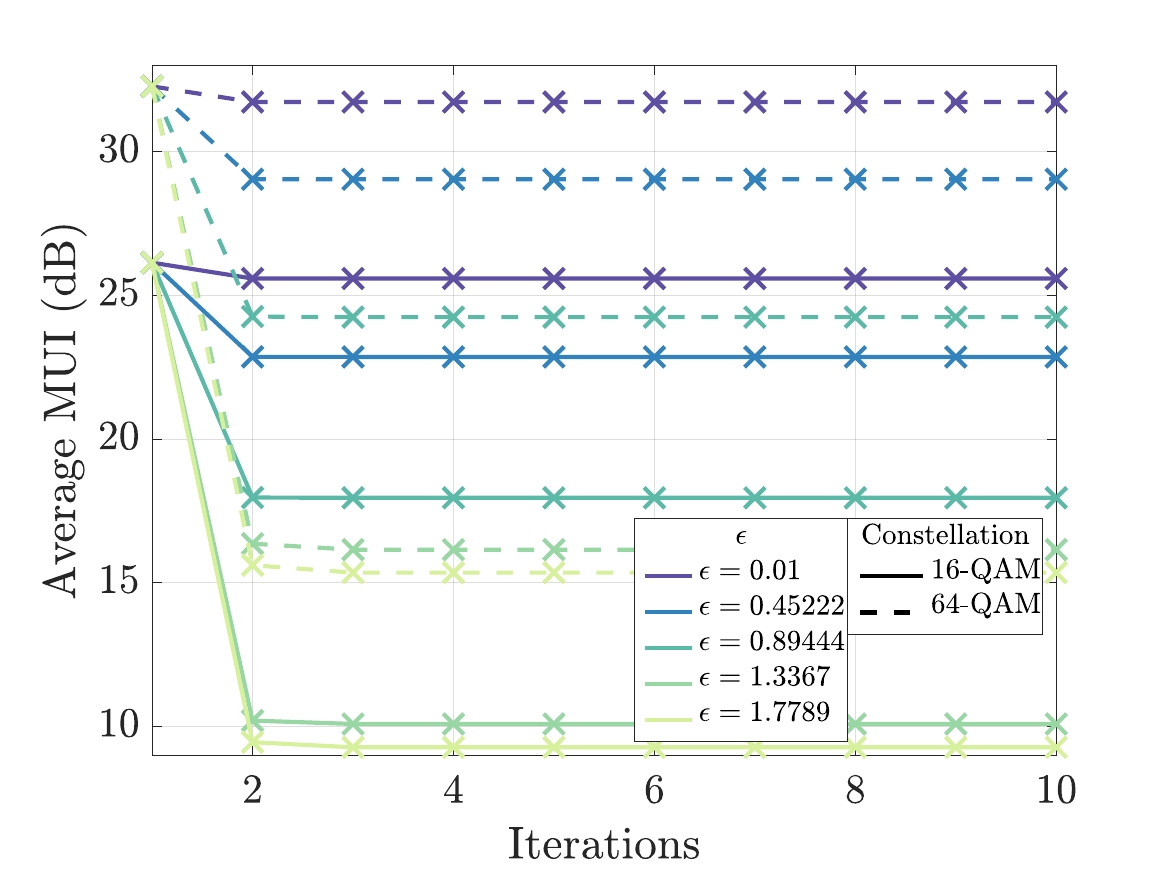}
 \caption{Average MUI over optimization iterations. The same parameters are set as that of Fig. \ref{fig:SINR-iter}.}
 \label{fig:MUI-iter}
\end{figure}

\begin{figure}[t]
 \centering
 \includegraphics[width=0.75\linewidth]{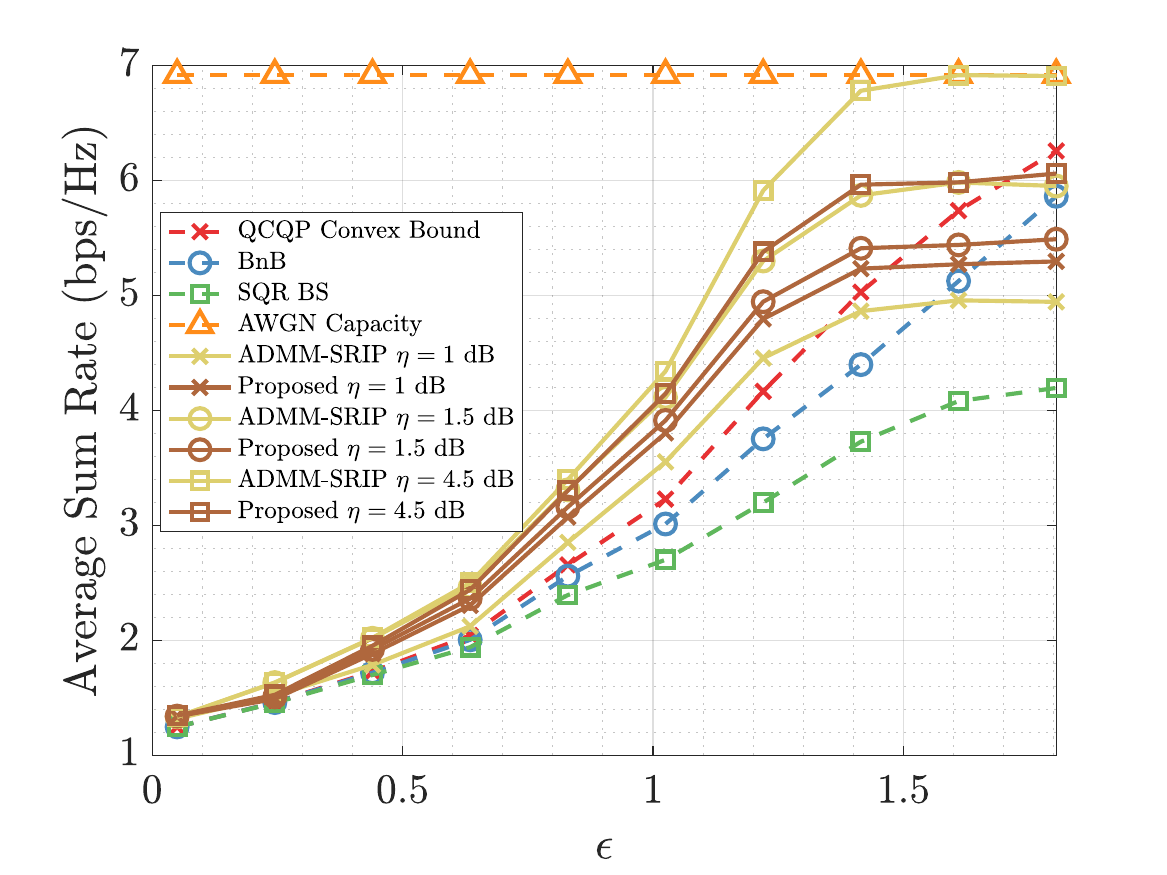}
 \caption{\textcolor{black}{The performance of the proposed waveform in terms of average sum rate vs. similarity rate. The number of Tx antennas is $N_T = 12$, the signal length $L = 7$, and the number of communication users $P =  4$ users.}
}
 \label{fig:benchmarking}
\end{figure}


\begin{figure*}[!t]
\centering
\subfloat[QPSK ($\eta =  6\dB$)]{\includegraphics[height=1.25in,width=1.66in]{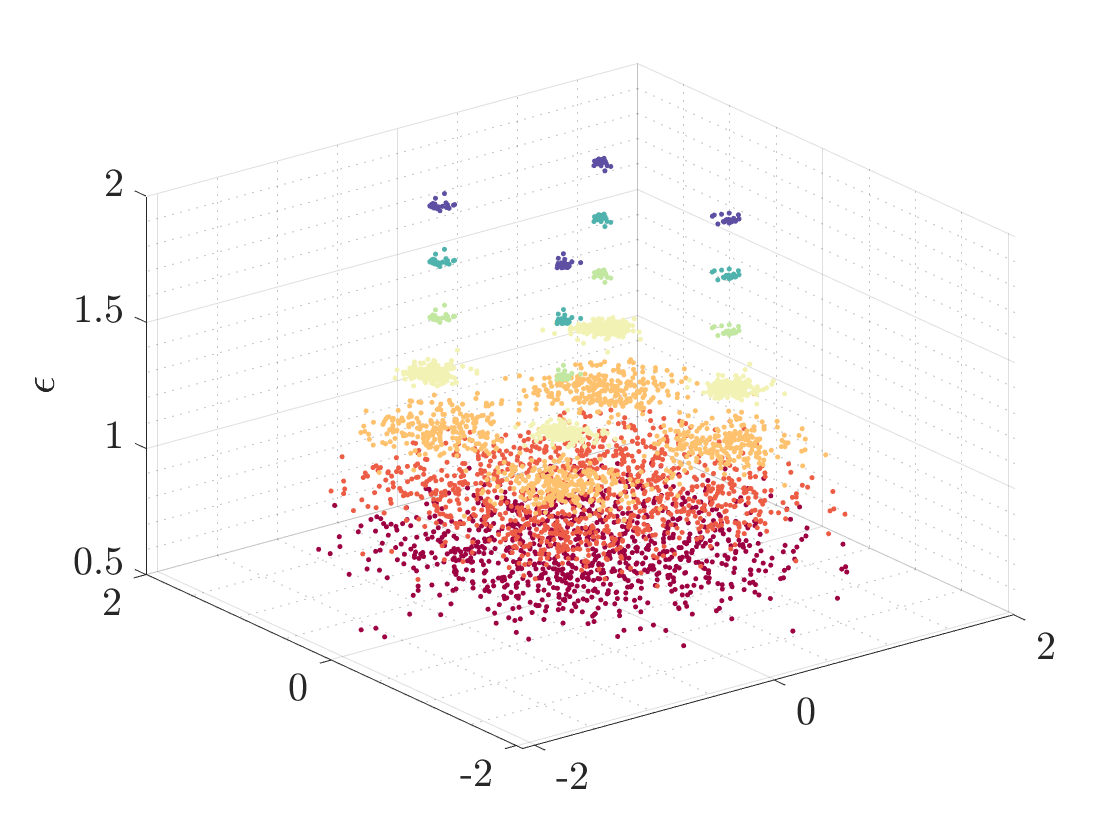} 
\label{fig:constellation-01}}
\hfil
\subfloat[16-PSK ($\eta =  6\dB$)]{\includegraphics[height=1.25in,width=1.66in]{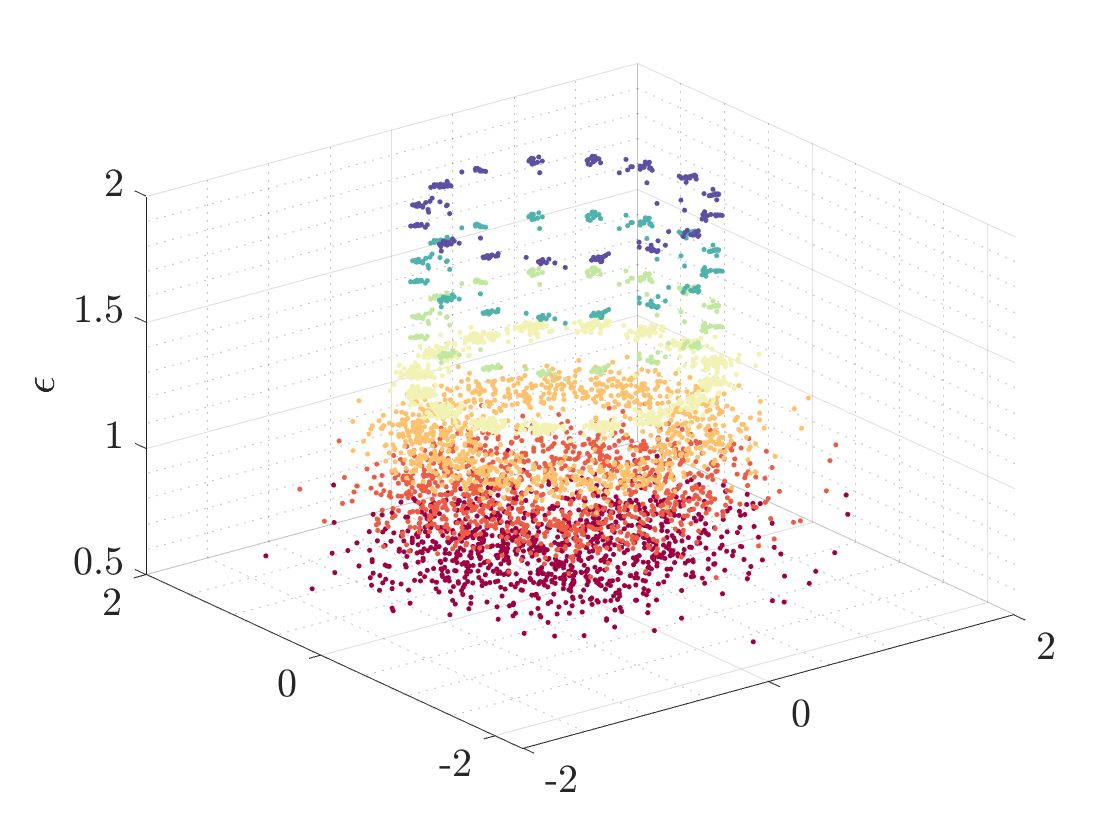}  
\label{fig:constellation-02}}
\subfloat[QPSK ($\epsilon =  2$)]{\includegraphics[height=1.25in,width=1.66in]{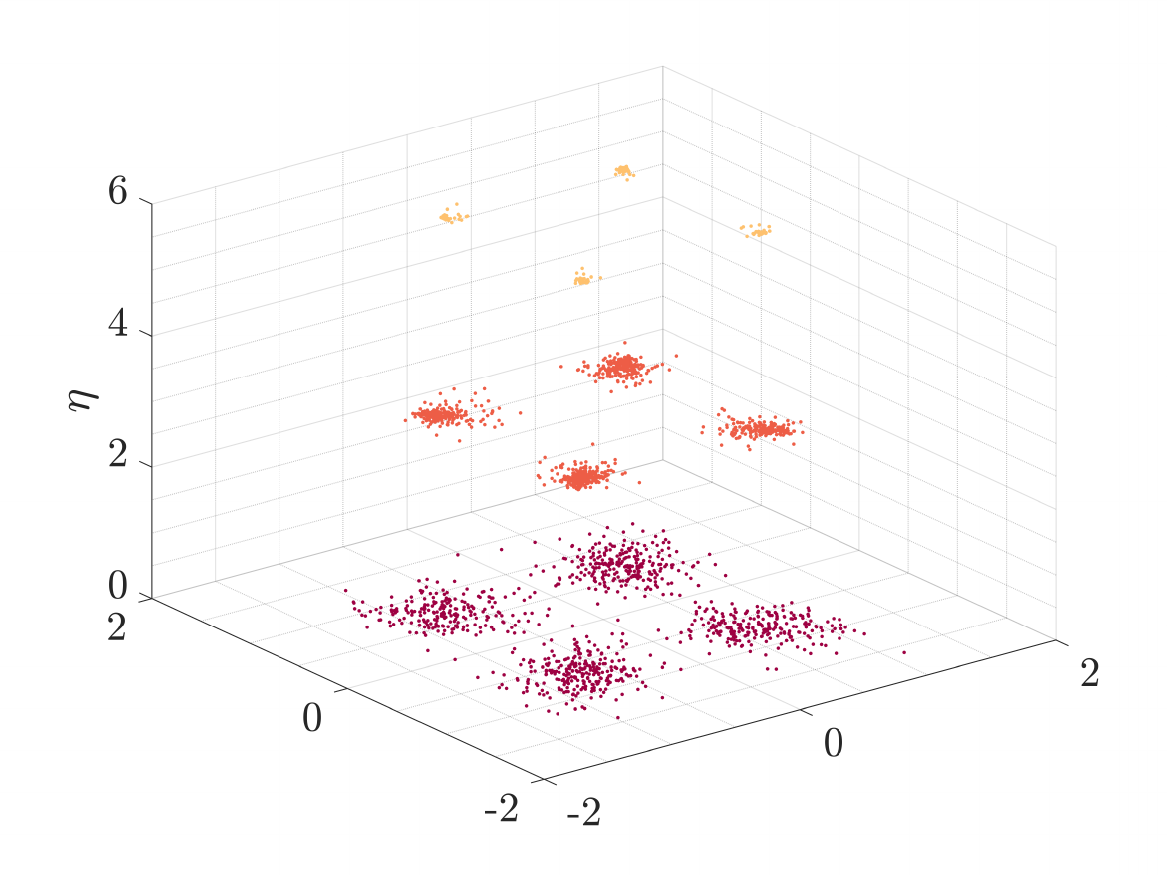}  
\label{fig:constellation-03}}
\subfloat[16-PSK ($\epsilon =  2$)]{\includegraphics[height=1.25in,width=1.66in]{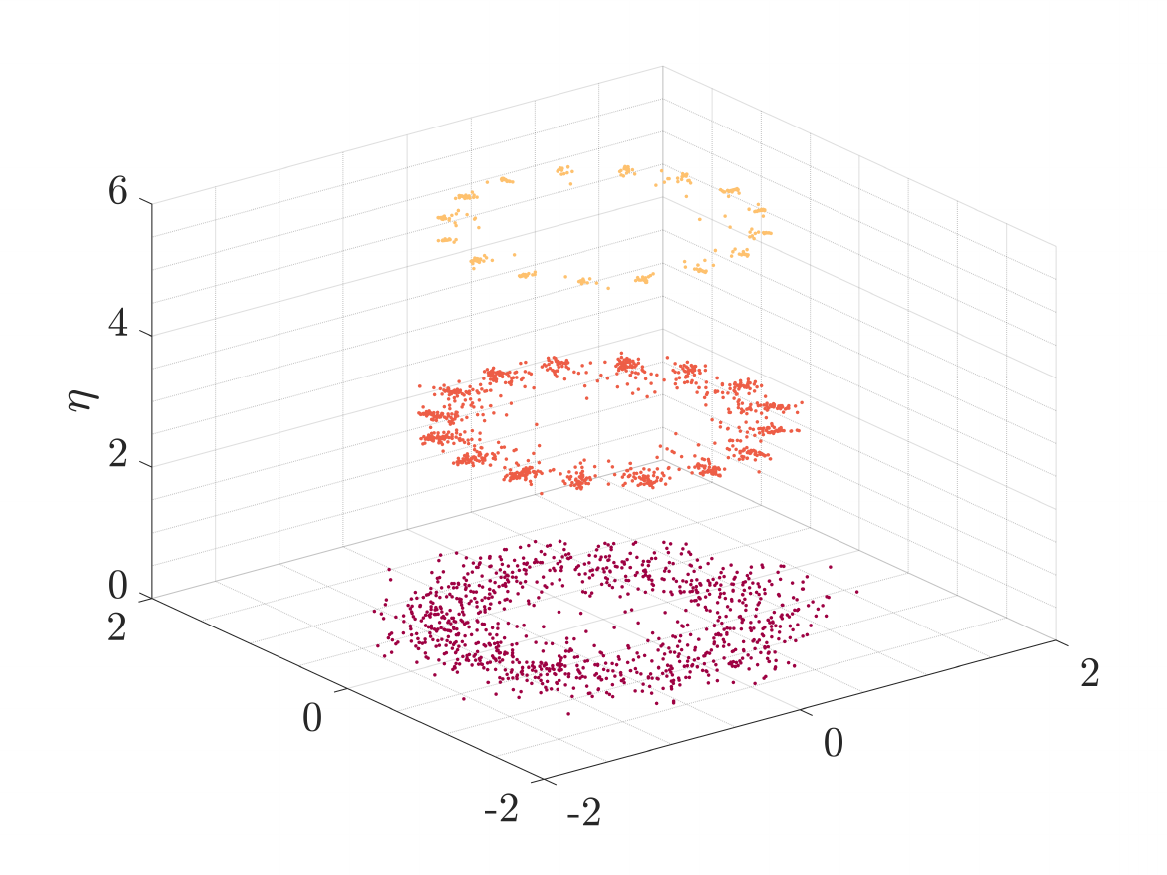}  
\label{fig:constellation-04}}
\caption{\textcolor{black}{\ac{DRIP} constellations over varying $\epsilon$ \& $\eta$, where $N_T= 12$, $L = 7$, $P = 5$. The x- and y- axis represent in-phase/quadrature values of $\pmb{HX}$, where $\pmb{X}$ is obtained through \ac{DRIP} (\textbf{Algorithm 1}). The captions of sub-figures indicate the constellations specified by $\pmb{S}$.}}
\label{fig:constellation}
\end{figure*}

\begin{figure}[t]
 \centering
 \includegraphics[width=0.75\linewidth]{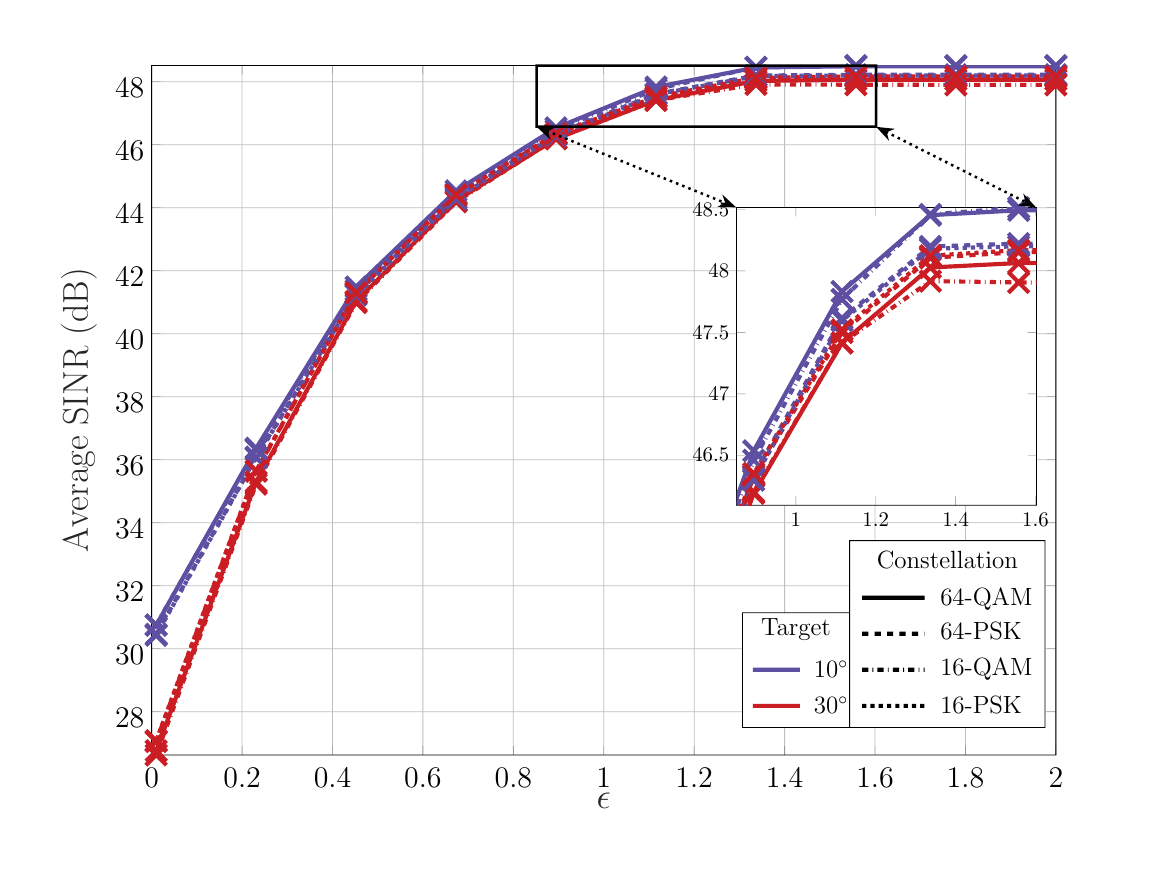}
 \caption{Trade-off within sensing between target SINR and $\epsilon$. 
 \textcolor{black}{The targets are assumed to be at equal distances from the BS.}
 The same parameters are set as that of Fig. \ref{fig:SINR-iter}.}
 \label{fig:SINR-eps}
\end{figure}

\textcolor{black}{\input{actions/benchmark-reference}} 
\begin{figure}[t]
	\centering
\includegraphics[width=0.75\linewidth]{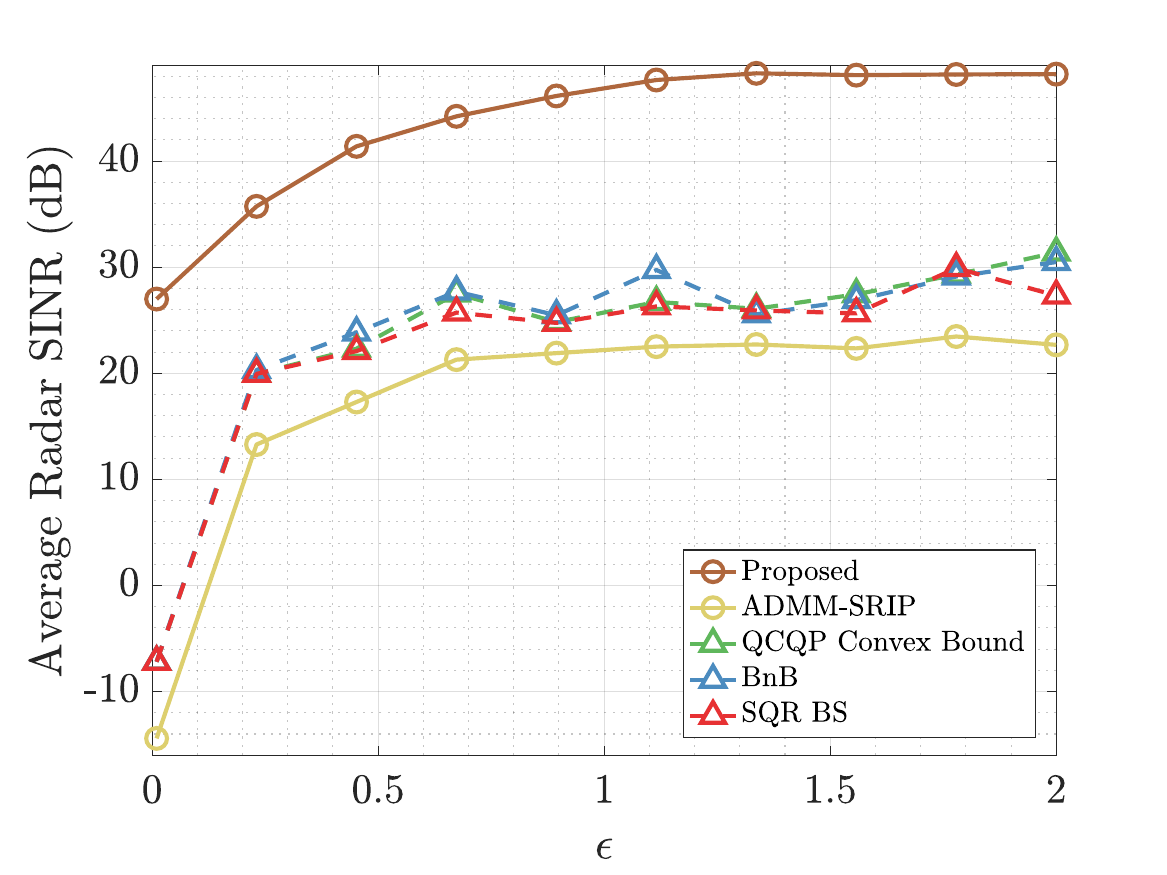}
 \caption{\textcolor{black}{Comparison of the average radar \ac{SINR} against $\epsilon$ for DRIP method against benchmarks.}}
 \label{fig:RadarSINR_Benchmark}
\end{figure}

\begin{figure*}[!t]
\centering
\subfloat[Targets at $\theta_1 = 10^{\circ}$ \text{ and } $\theta_2 = 30 ^{\circ}$]{\includegraphics[height=2in,width=2.66in]{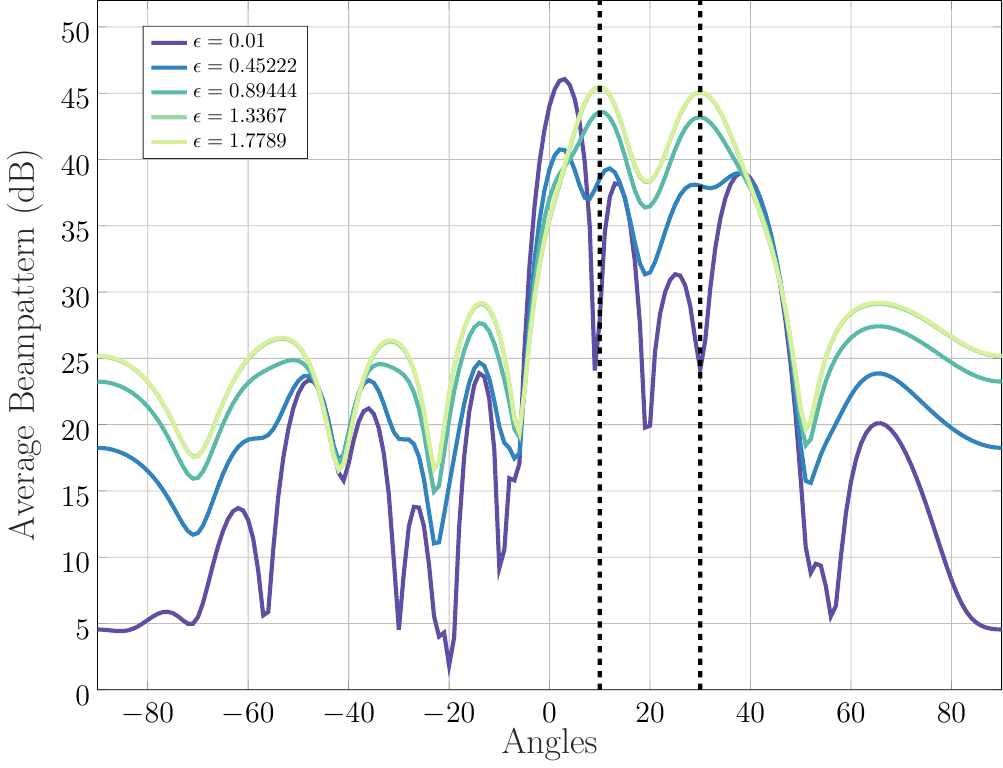} 
\label{fig:beampattern-01}}
\hfil
\subfloat[Targets at $\theta_1 = -40^{\circ}$ \text{ and } $\theta_2 = 20 ^{\circ}$]{\includegraphics[height=2in,width=2.66in]{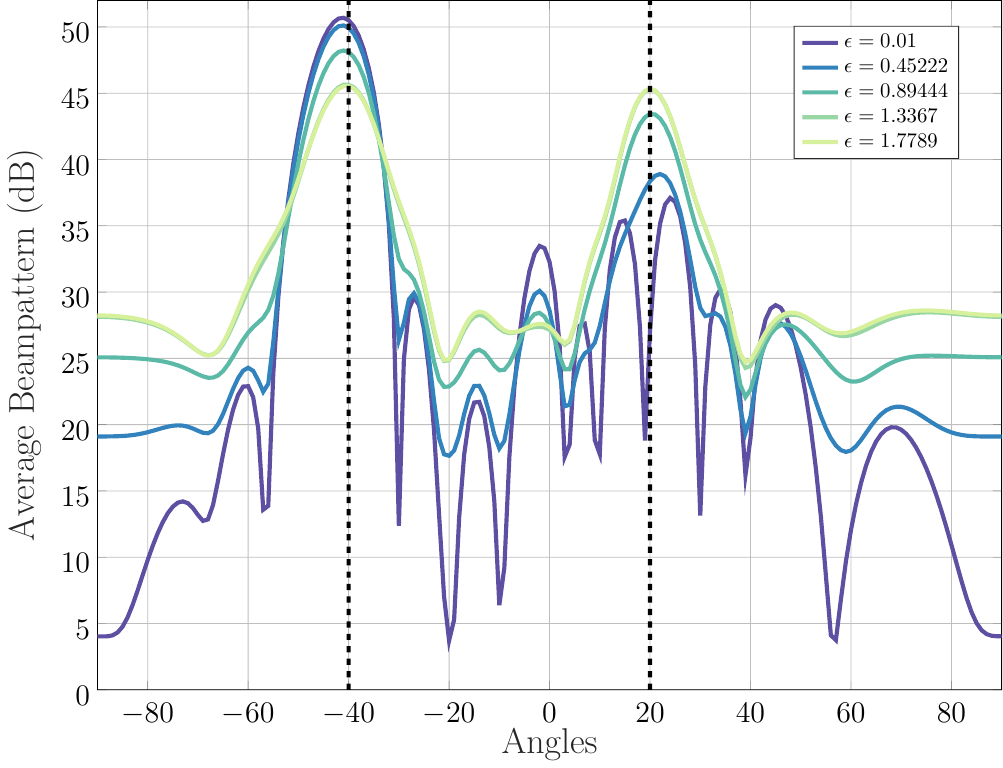}  
\label{fig:beampattern-02}}
\caption{Resulting beampatterns for different 2-target configurations. 
The number of transmit antennas is $N_T = 12$, the signal length $L = 7$, the number of communication users $P = 4$, $\eta = 2.5 \dB$ and the utilized constellation was a $64$-\ac{QAM}. Dotted vertical lines correspond to the true target locations.}
\label{fig:beampattern}
\end{figure*}

\begin{figure*}[!t]
\centering
\subfloat[Fixed $\epsilon = 0.01$]{\includegraphics[height=2in,width=2.66in]{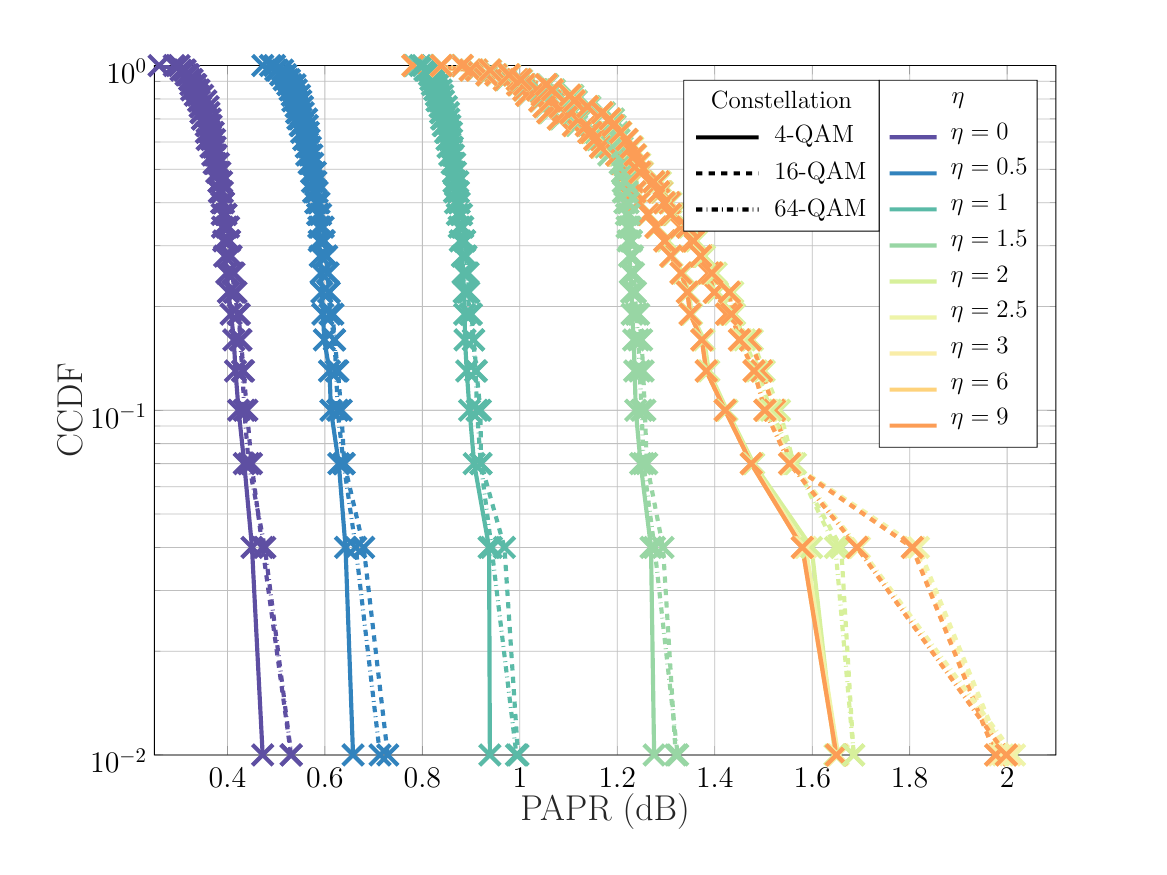} 
\label{fig:ccdf-01}}
\hfil
\subfloat[Fixed Constellation = 16-QAM]{\includegraphics[height=2in,width=2.66in]{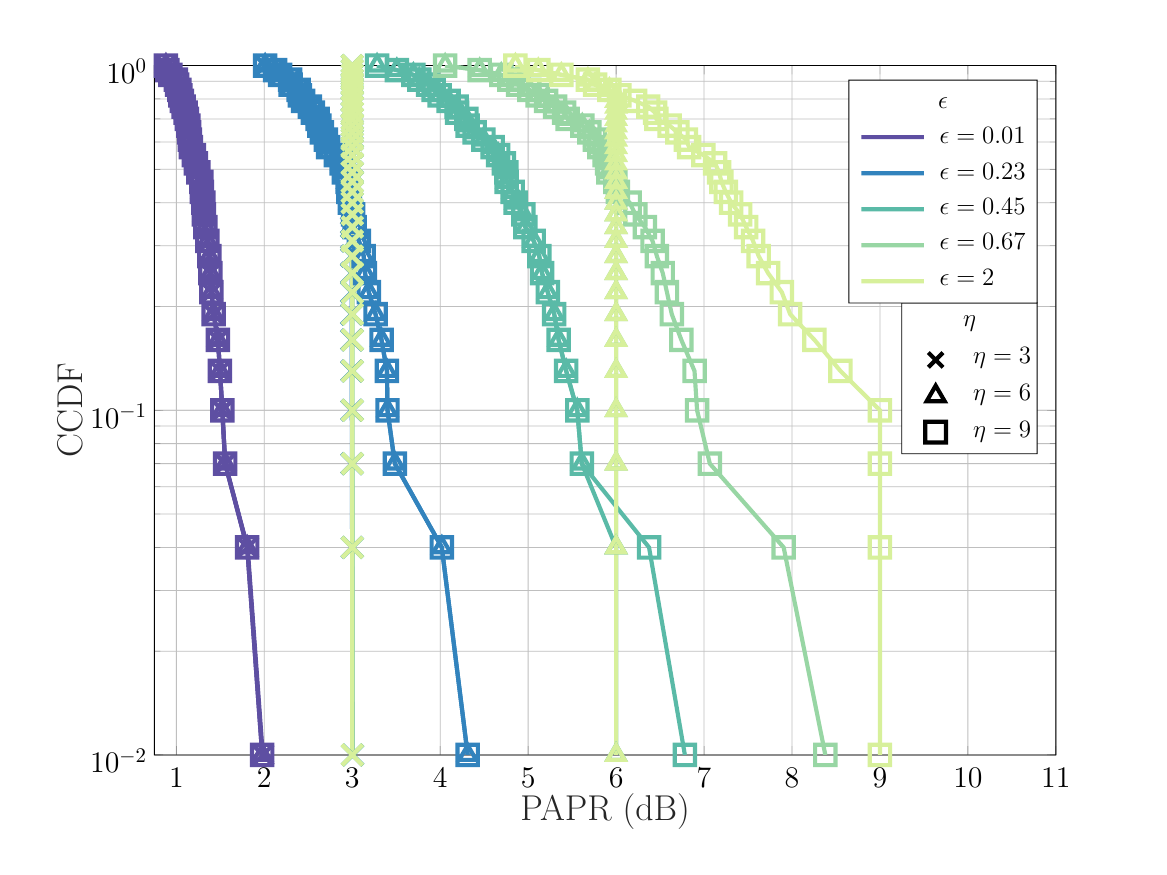}  
\label{fig:ccdf-02}}
\caption{\ac{CCDF} of \ac{PAPR} for two different sets of testing parameters. The parameters $N_T$, $L$ and $P$ are the same as in Fig. \ref{fig:beampattern}.}
\label{fig:ccdf}
\end{figure*}

\begin{figure}[t]
 \centering
 \includegraphics[width=0.75\linewidth]{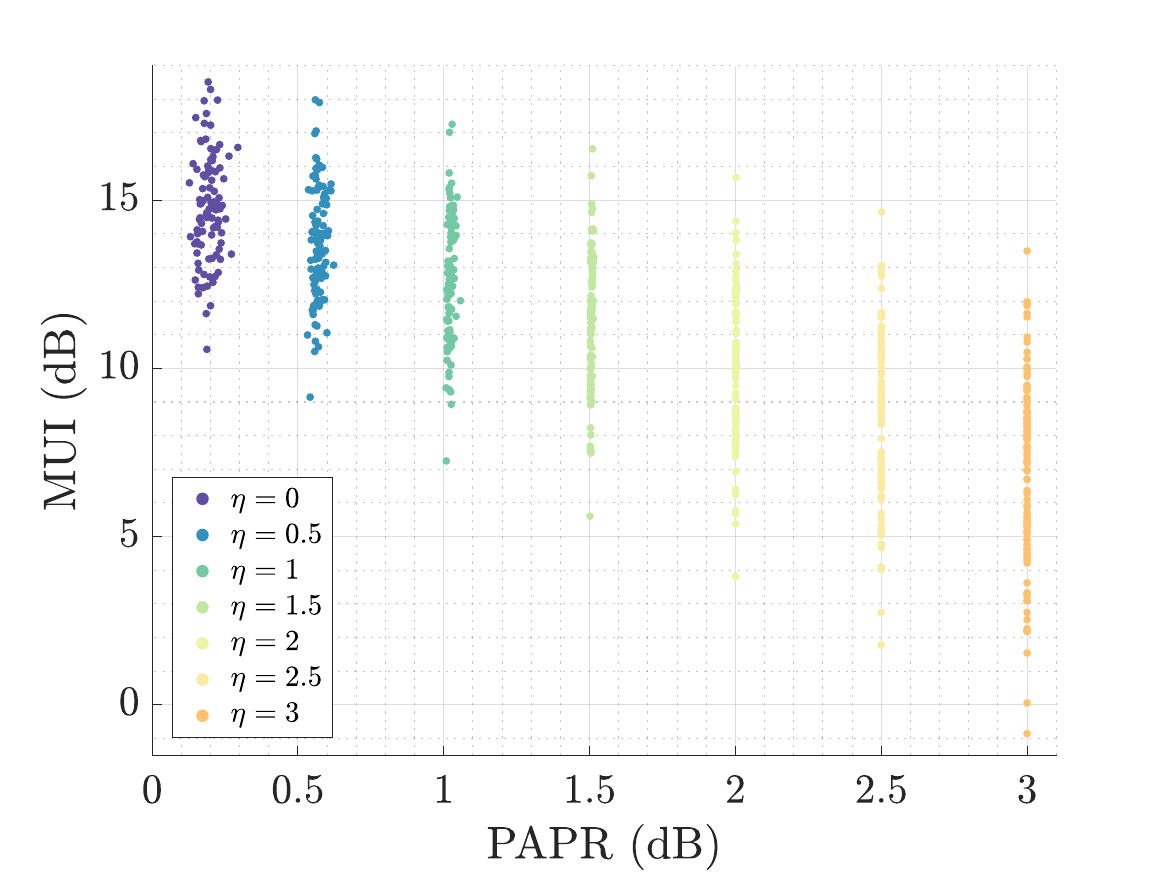}
 \caption{\textcolor{black}{Instantaneous MUI vs instantaneous PAPR for $N_T =12$, $N_R = 7$, $L =7$, $P=4$, $\rho=10$, $\sigma_v^2 = 0.01$ for $16$-QAM.}}
 \label{fig:inst-MUI-PAPR}
\end{figure}

\begin{figure}[t]
 \centering
 \includegraphics[width=0.75\linewidth]{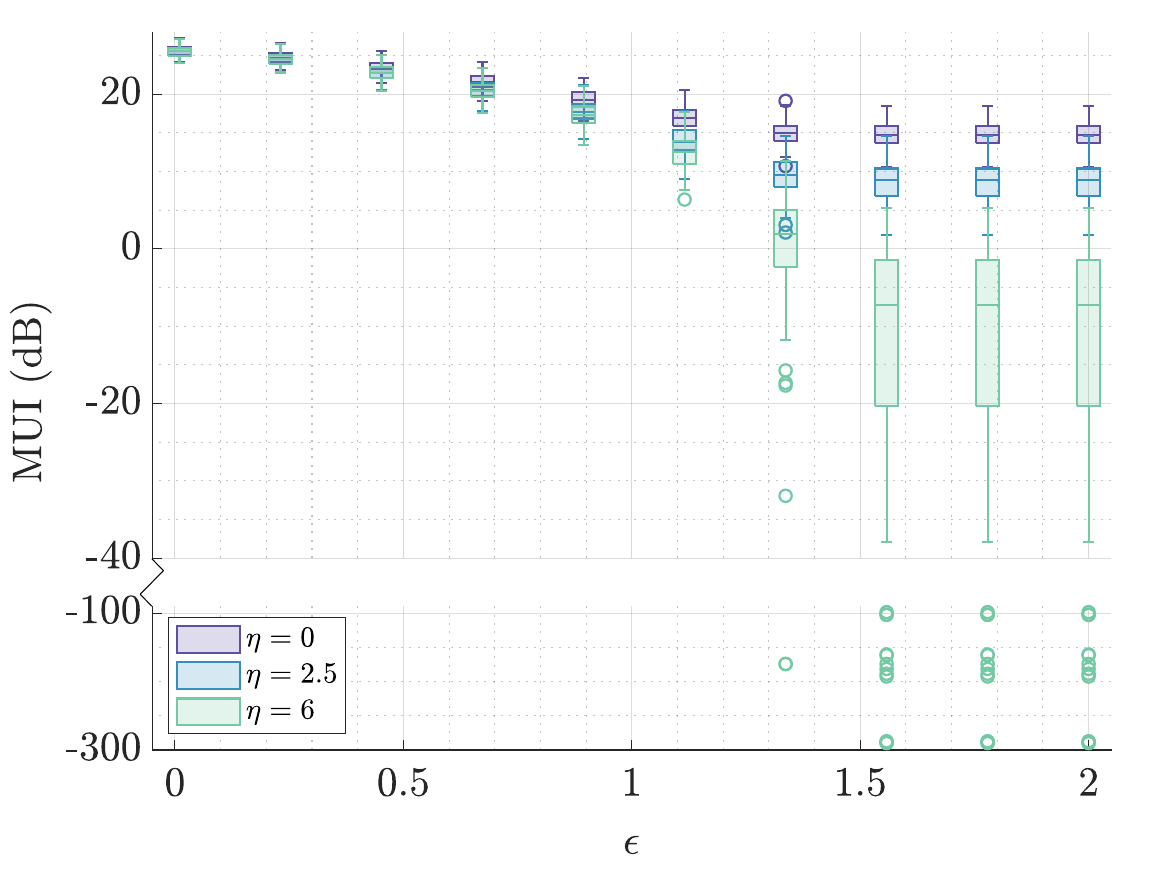}
 \caption{MUI distribution with respect to $\epsilon$. The number of Tx antennas is 12, the signal length $L = 7$, the number of communication users $P = 4$ , and the modulation used is 16-QAM. 
}
 \label{fig:MUI-eps-box}
\end{figure}

\begin{figure}[t]	
 \centering
 \includegraphics[width=0.75\linewidth]{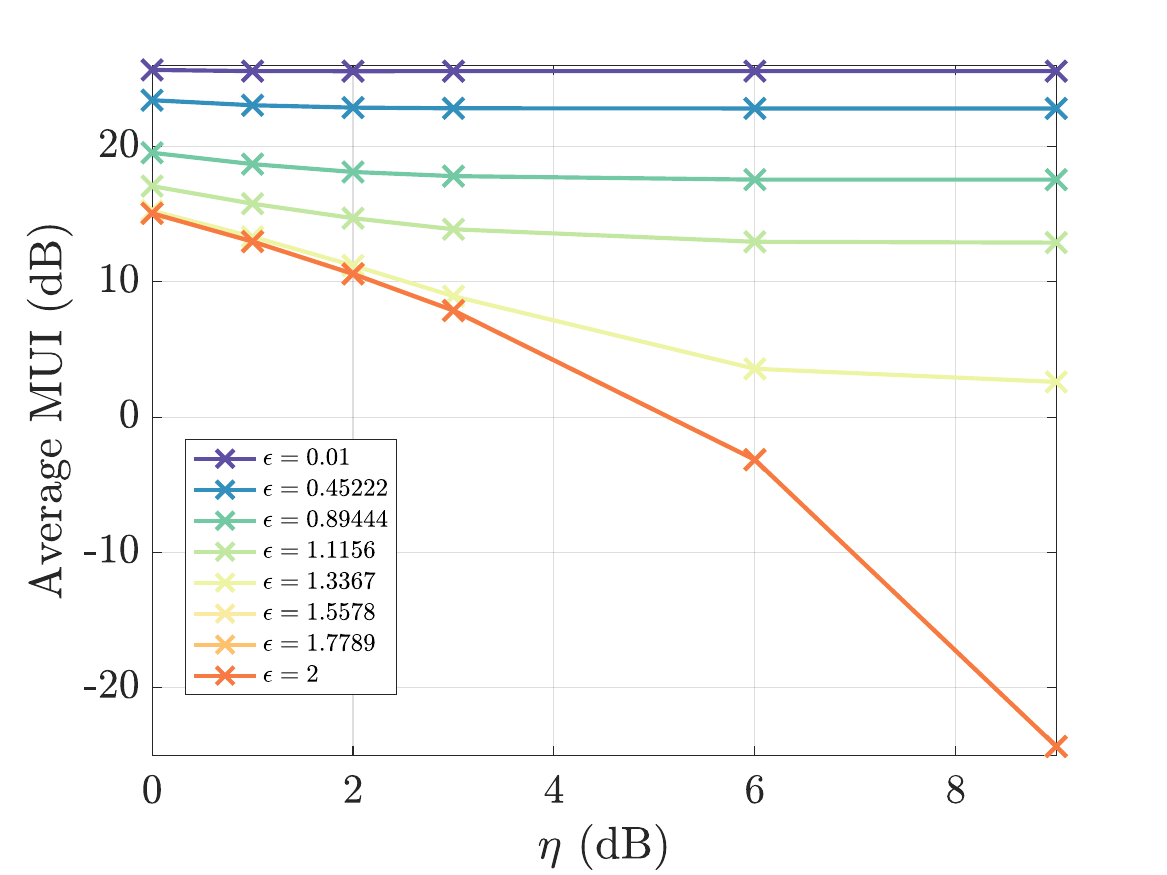}
 \caption{Average MUI with respect to $\eta$. 
The number of Tx antennas is 12, the signal length $L = 7$, the number of communication users $P = 4$ , and the modulation used is $16$-\ac{QAM}. }
 \label{fig:MUI-eta}
 
\end{figure}

\begin{figure}[t]
 \centering
 \includegraphics[width=0.75\linewidth]{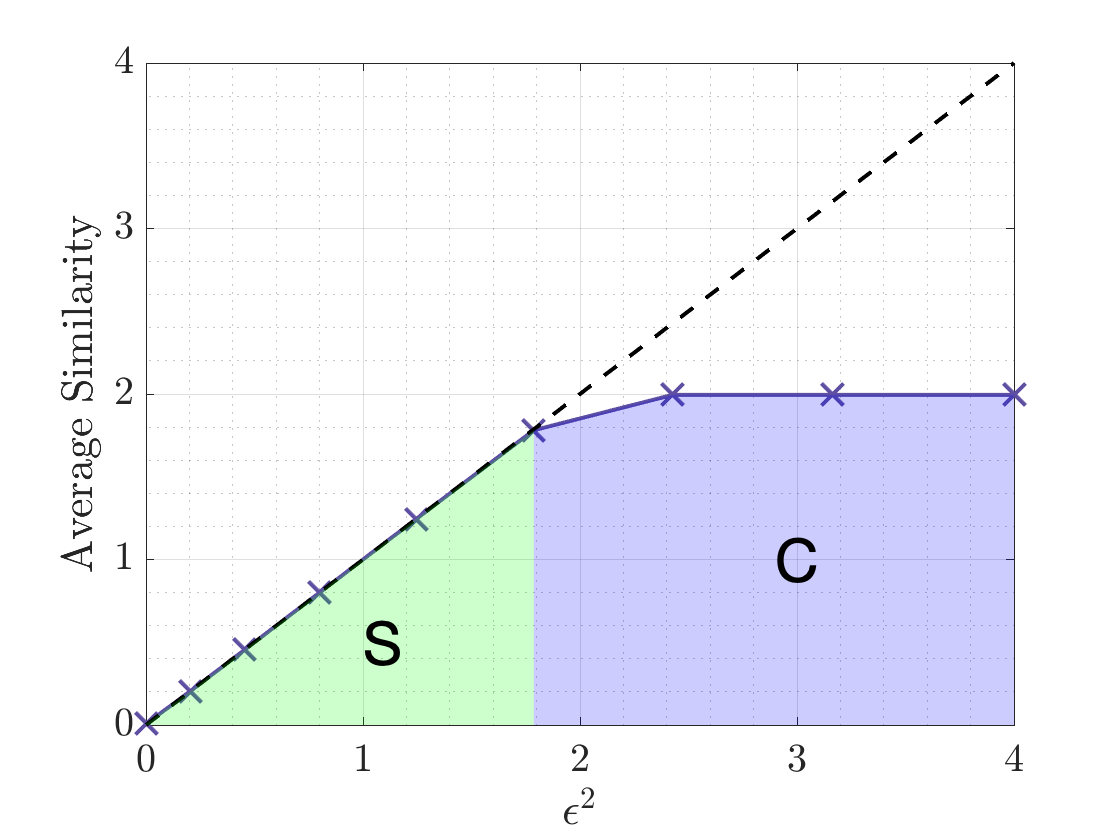}
 \caption{Average empirical similarity with the radar chirp with respect to $\epsilon$. The curve shown is get from all variations of $\eta$, $\bar{g}_q$, and constellations.}
 \label{fig:s-and-c}
\end{figure}

\paragraph{Convergence behavior of \ac{BCCD} for \ac{DRIP}}
In Fig. \ref{fig:SINR-iter}, the radar \ac{SINR} performance is studied over \textcolor{black}{the number of outer iterations} of \textbf{Algorithm \ref{alg:alg1}}, i.e. the \ac{BCCD} algorithm for \ac{DRIP}.
In this simulation, the following parameters are used: $N_T = 12$, $L = 7$, $P = 4$, and $\eta = 2.5 \dB$. Two targets are set in the scene at $\theta_1 = 10^\circ$ and $\theta_2 = 30^\circ$.
Here, the average \ac{SINR} is shown for the first target.
For all iterations, the radar \ac{SINR} measured at this target is above the $20 \dB$ constraint imposed by $\bar{g}_1 = \bar{g}_2 = 20\dB$. We can see that the \ac{SINR} increases steadily before reaching a converging ceiling typically at the third iteration. We can see that in this case, the \ac{SINR} at this target increases with $\epsilon$.
This is because a loose similarity constraint allows for a more optimistic \ac{SINR} beamforming towards the targets, making the signal power more concentrated to the different target directions, hence exceeding the thresholds specified by $\bar{g}_1$ and $\bar{g}_2$. This can also be seen in Fig. \ref{fig:beampattern}. 
Also, an increase in the \ac{QAM} size slightly gives a less average SINR at convergence.

In Fig. \ref{fig:MUI-iter}, we study the convergence behaviour of the communication \ac{MUI} cost with \textcolor{black}{the number of outer iterations}. 
In this simulation, the same parameters are set as that of Fig. \ref{fig:SINR-iter}.
Also, with a looser $\epsilon$ constraint, we can see a decrease in the \ac{MUI} floor, which implies higher communication rate. With the constellation increasing from $16$-QAM to $64$-QAM, we observe a rough increase of about $3 \dB$ \textcolor{black}{to $6 \dB$} over all \ac{MUI} values. 
This is because more symbols are being transmitted under a fixed transmit power, which translates into increased \ac{MUI}.
In all cases, we observe convergence in few iterations.

\paragraph{\ac{DRIP} rate-similarity tradeoffs}
In Fig. \ref{fig:benchmarking}, we study the trade-off between average communication sum rate and similarity constraint with the radar chirp. 
In this simulation, the following parameters are used: the number of transmit antennas is set to $N_T = 4$, the number of communication users is fixed to $P = 2$, and the utilized modulation scheme is \ac{QPSK}.
The waveform length is set to $L = 7$ samples.

\textcolor{black}{\input{actions/benchmark-reference}}
\textit{A very important note herein is that all other methods do not incorporate beamforming capabilities and only optimize the transmit waveform signal over the temporal domain.}
\textcolor{black}{\input{actions/sum-rate-computation}}
Note that when the radar similarity rate is very stringent ($\epsilon \rightarrow 0$), then all methods exhibit a low average sum rate. For example, at $\epsilon = 0.05$, all methods operate at a similar sum rate of $1.25 \bpspHz$.
As $\epsilon$ increases, i.e. when the radar constraint becomes less demanding, we have interesting observations.
When a low \ac{PAPR} \ac{ISAC} signal is desired, such as the case of $\eta = 1 \dB$, the proposed method, i.e. \ac{DRIP}, performs better than that in \cite{10061453}. 
In particular, it can achieve better communication performance under the same sensing similarity constraint as compared to \ac{SRIP}, the \ac{BnB} method, and the \ac{SQRBS} method.
More specifically, when $\epsilon = 1.025$ which can be thought of as midway between radar similarity and communications, 
\ac{DRIP} attains an average sum rate of $3.8 \bpspHz$, whereas 
\ac{SRIP} achieves $3.5 \bpspHz$. Moreover,
\ac{BnB} and \ac{SQRBS} achieve $3 \bpspHz$ and $2.7 \bpspHz$, respectively.
This is because, for low \ac{PAPR}, the dual beam-similarity constraints can better assist for radar tasks leaving more efficient resources for \ac{MUI} communication performance.
For a loose $\epsilon$, \ac{DRIP} can achieve an average sum rate of up to $6\bpspHz$.   
However, at higher $\eta$ values, \ac{SRIP} outperforms \ac{DRIP} and can achieve the \ac{AWGN} capacity.
For example, at $\epsilon = 1.225$, \ac{SRIP} outperforms the proposed method by roughly $0.6 \bpspHz$ at $\eta = 4.5$ dB and by roughly $0.5\bpspHz$ at $\eta = 1.5$ dB.
\textcolor{black}{\input{actions/crossover-eta}}


\paragraph{\textcolor{black}{($\epsilon,\eta$)-dependent \ac{DRIP} constellations}}
In Fig. \ref{fig:constellation}, the constellation of \ac{DRIP} waveforms are characterized over multiple $\epsilon$ values. 
In this simulation, the following parameters are used: the number of Tx antennas is $N_T = 12$, the signal length $L = 7$, the number of communication users is $P = 5$, and $\eta =  6 \dB$.
\textcolor{black}{\input{actions/epsilon-eta-papr}}

\paragraph{Dual sensing trade-offs}
In Fig. \ref{fig:SINR-eps}, we study the trade-off within sensing between prioritizing target \ac{SINR} and similarity with the chirp. 
The same parameters are set as that of Fig. \ref{fig:SINR-iter}.
We notice that an increase in $\epsilon$ contributes to an increase in average radar \ac{SINR} across all the involved targets of interest.
In particular, an increase in radar \ac{SINR} necessitates a sacrifice in similarity rate for a fixed $\bar{g}_q$.  
This is because, for fixed power, relaxing the temporal domain constraint over the radar similarity chirp leaves more opportunity for an increase in target \ac{SINR}.
In addition, the \ac{PSK} constellations can give an even \textcolor{black}{better} \ac{SINR} performance across targets than \ac{QAM}. For example, at $\epsilon = 1.3367$, $16$-\ac{QAM} gives an \ac{SINR} disparity between targets of about $0.6 \dB$, while $16$-\ac{PSK} gives about $0.1 \dB$.
\textcolor{black}{\input{actions/gap-10-30}}

\textcolor{black}{\input{actions/DRIP-compare-SINR}}

\vspace{-0.3cm}
\paragraph{\ac{DRIP} beampatterns}
In Fig. \ref{fig:beampattern}, we plot the beampattern of the proposed \ac{DRIP} waveforms, which is obtained by steering the \ac{SINR} over  angles from $-90^\circ$ to $90^\circ$. 
In this simulation, the same parameters are set as that of Fig. \ref{fig:SINR-iter}, except for $\eta = 2.5 \dB$ and the utilized constellation was a $64$-\ac{QAM}.
When we lower the $\epsilon$, thus making the waveform more similar to a radar chirp, we can achieve better sidelobe performances but very ambiguous peaks, meaning that the beamforming is not ideal. 
As $\epsilon$ increases, the \ac{DRIP} beampattern tends to point towards the targets of interest , which is translated to sharp peaks towards the targets of interest. 
By comparing Fig. \ref{fig:beampattern-01} with Fig. \ref{fig:beampattern-02}, we can also observe that when the targets are further away, \textit{clearer peaks can be achieved even at lower $\epsilon$ values, which translates into improved \ac{DRIP} spatial resolution}. For example, at $\epsilon = 0.45222$, Fig. \ref{fig:beampattern-02} shows two clear peaks of about $48 \dB$ and $43 \dB$ at the desired target locations while its counterpart, Fig. \ref{fig:beampattern-01}. only shows an ambiguous plateau of around $37 \dB$ between $0^\circ$ and $40^\circ$.
\textcolor{black}{\input{actions/delay-doppler-angular}}

\paragraph{\ac{DRIP} \ac{PAPR} \ac{CCDF} statistics}
In Fig. \ref{fig:ccdf}, we analyze the complementary cumulative distribution function (CCDF) of the PAPR values under different $\eta$.
The parameters $N_T$, $L$ and $P$ are the same as in Fig. \ref{fig:beampattern}.
Fig. \ref{fig:ccdf-01} shows how the \ac{CCDF} distribution of the \ac{PAPR} distribution varies with different $\eta$, at a fixed $\epsilon$, which is fixed to $\epsilon = 0.01$.
As the $\eta$ constraint loosens, the distribution of \ac{PAPR} spreads more to higher values as expected. 
In fact, the \ac{CCDF} of \ac{PAPR}s converges to a certain distribution when $\eta$ grows large. 
This is because the actual \ac{PAPR}s of the \ac{DRIP} waveforms live beyond these large $\eta$ values, hence after a certain $\eta$, it is "as if" the \ac{PAPR} constraint becomes inactive.
For instance, at $\eta \geq 3 \dB$, we observe similar distributions across all $\eta$ values. 
In addition, the \ac{PAPR} distribution tends to higher values when the constellation size is increased.
Fig. \ref{fig:ccdf-02} shows how the \ac{PAPR} distribution varies with $\epsilon$, for a fixed constellation. 
In this simulation, we fix the modulation to $16$-\ac{QAM}. As expected, a looser $\epsilon$ promotes \ac{DRIP} waveforms of higher \ac{PAPR} levels. 
Also, we can see that at high $\epsilon$, the PAPR is strongly restricted by $\eta$. This is because \ac{DRIP} waveforms become primarily constrained by the \ac{PAPR} constraint instead of the similarity constraint.
\textcolor{black}{\input{actions/eta-epsilon-papr-2}}
\textcolor{black}{\input{actions/eta-papr-mui}}

\begin{figure}[t]
	\centering
\includegraphics[width=0.75\linewidth]{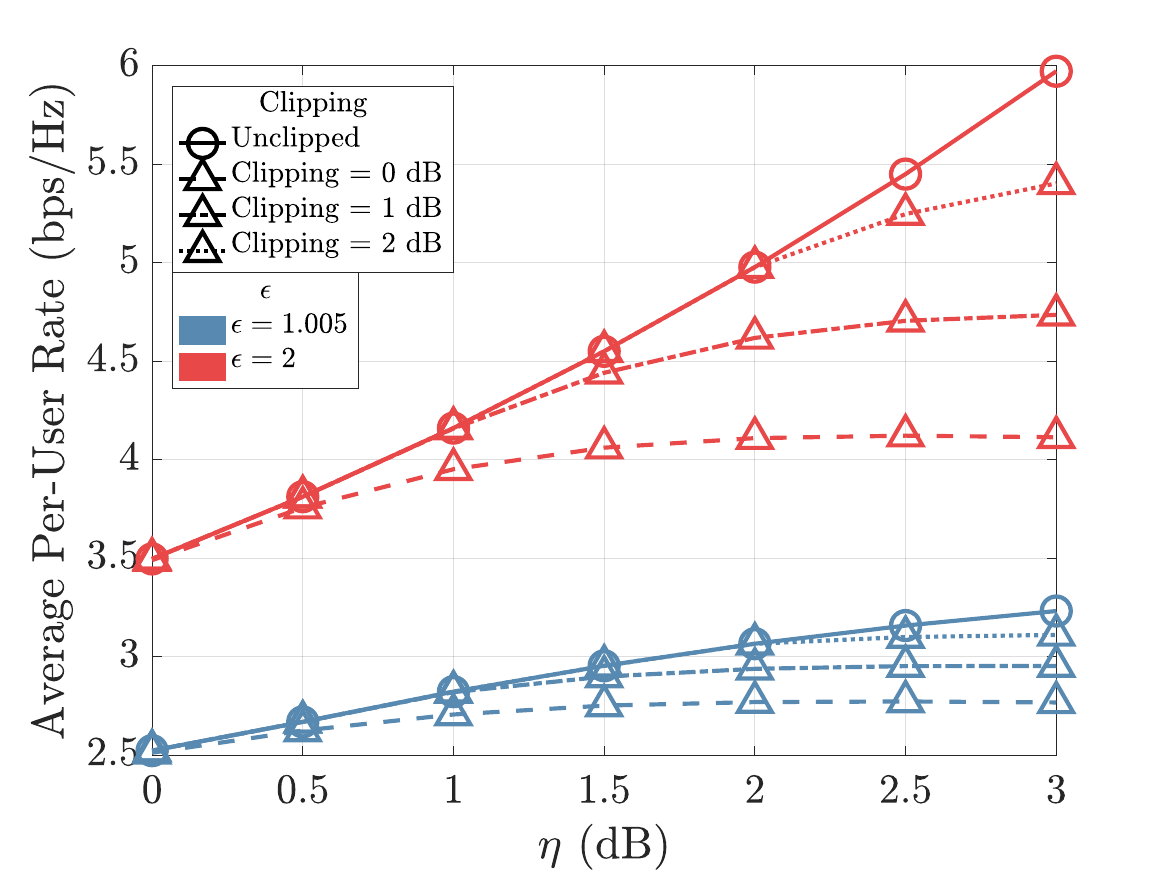}
 \caption{\textcolor{black}{Average per-user rate vs. $\eta$ for different \ac{HPA} clipping levels. The simulation parameters are $N_T = 12$, $N_R = 7$, $L = 7$, $P = 4$, $\rho = 10$, $\sigma_v=0.01$ and $16$-QAM.}}
 \label{fig:HPA}
\end{figure}

\paragraph{\ac{DRIP} \ac{MUI} variability with similarity}
In Fig. \ref{fig:MUI-eps-box}, we study the distribution of the MUI, which is directly related to the sum rate for a fixed constellation, with respect to $\epsilon$. In this simulation, the following parameters are used: the number of Tx antennas is $12$, the signal length $L = 7$, the number of communication users $P = 4$, and the modulation used is 16-QAM. 
Lower median \ac{MUI} values with higher variations of \ac{MUI} are observed at higher values of $\epsilon$.
For example, when $\epsilon = 1.33$, 
the median \ac{MUI} of \ac{DRIP} waveforms at $\eta = 6 \dB$ is \textcolor{black}{$1.87\dB$} with the lower and upper whiskers corresponding to \textcolor{black}{$-11.77 \dB$ and $11.31 \dB$}, respectively, which is an \ac{MUI} gap of \textcolor{black}{$23\dB$}. 
When $\epsilon \geq 1.55$, the median \ac{MUI} drastically drops to \textcolor{black}{$-7.2\dB$} with an \ac{MUI} gap becomes \textcolor{black}{$43\dB$}.
Generally speaking, as $\epsilon$ increases, the median \ac{MUI} values of \ac{DRIP} waveforms decreases with an increase in \ac{MUI} whisker gaps, i.e. more variability in \ac{MUI}.

In Fig. \ref{fig:MUI-eta}, we illustrate the average \ac{MUI} behavior, which has an impact on the communication rate, with increasing $\eta$ and different $\epsilon$ values.
In this simulation, the following parameters are used: the number of Tx antennas is $N_T = 12$, the signal length $L = 7$, the number of communication users $P = 4$ users, and the modulation used is $16$-\ac{QAM}. 
The higher the $\eta$, the less the \ac{MUI} cost. This is because it allows for larger \ac{PAPR}, making it closer to an ideal communication waveform. 
We can see the trade off between $\eta$ and $\epsilon$ along with their impact on \ac{MUI}. 
For low $\epsilon$ values (i.e. $\epsilon \leq 1.11$), the prioritizing of similarity manifests as the ``floor''  at high \textcolor{black}{$\eta$}, when the PAPR constraint is loose. For higher $\epsilon$, the curves coincide.
In particular, for $\epsilon \geq 1.55$, the behavior of \ac{MUI} vs. $\eta$ is the same, 
meaning that increasing $\epsilon$ beyond $1.55$ does not help much with obtaining a better communication rate. 
However, we can see that the \ac{MUI} floor (appearing only for lower $\epsilon$ values)  disappears, which is more favorable for communications. 
Hence, combined with discussions Fig. \ref{fig:s-and-c}, if we want to prioritize communications, we can mainly increase $\epsilon$ in the S-constrained region and $\eta$ in the C-constrained region. If we want to prioritize sensing similarity, which gives better space resolution of targets, we can decrease $\epsilon$ in the S-constrained region but the performance does not respond much to the tuning of any parameters in the C-constrained region.

\paragraph{Sensing and Communication Regions}
In Fig. \ref{fig:s-and-c}, we intend to illustrate the sensing (denoted as $\rm{S}$) and communication (denoted as $\rm{C}$) regions.
\textcolor{black}{\input{actions/average-similarity}}   
Specifically, the curve shown is obtained by varying a wide range of $\eta$ values spanning $0\dB$ to $9 \dB$, a wide range of $\bar{g}_q$ from $10 \dB$ and $20 \dB$, and different constellations of \ac{QPSK}, $16$-\ac{QAM}, $64$-\ac{QAM}, $16$-\ac{PSK}, and $64$-\ac{PSK}. 
It can be seen that the average empirical similarity is always constrained by $\epsilon$. 
For  $\epsilon^2 \leq 1.78$ values, the similarity measure changes with $\epsilon$ and attains the constraint with equality as the average similarity is on the straight line. 
This is because the \ac{DRIP} waveforms favor the similarity constraint, $\mathcal{B}_\epsilon(\pmb{x}_0)$, hence the reason we quantify this region as the $\rm{S}$-region.
When, $\epsilon^2 \geq 1.78$, the \ac{DRIP} waveforms favor the communication constellations and no longer adhere to the sensing chirp waveforms, so the waveforms enter the $\rm{C}$-region.

\vspace{-0.25cm}
\textcolor{black}{\input{actions/HPA-impact}}

\begin{figure}[t]
	\centering
\includegraphics[width=0.75\linewidth]{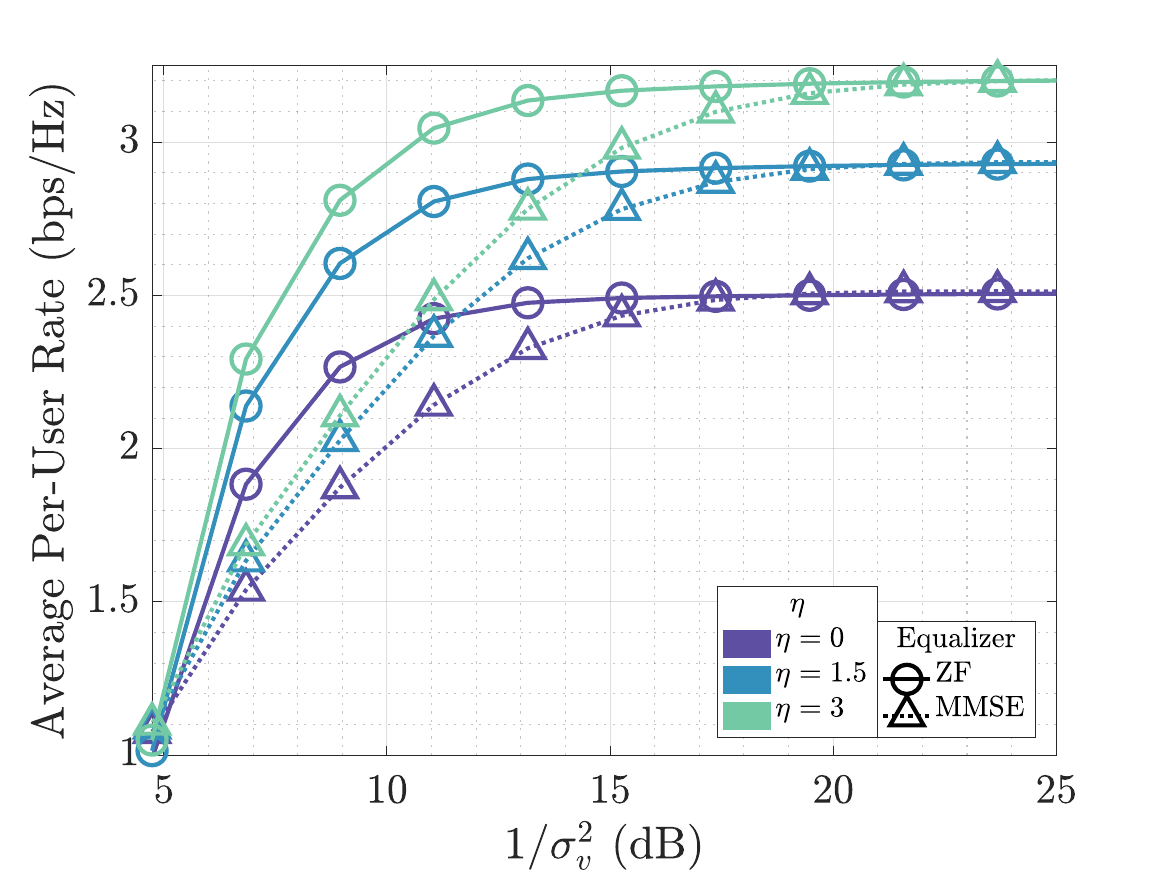}
 \caption{\textcolor{black}{Comparison of \ac{MMSE} vs \ac{ZF} approaches for $N_T = 12$, $N_R = 7$, $L = 7$, $P = 4$, $\rho = 10$, $\epsilon = 1$ and $16$-QAM.}}
 \label{fig:MMSE-vs-ZF}
\end{figure}
\vspace{-0.25cm}

\textcolor{black}{\input{actions/ZF-vs-MMSE}}

%% file: actions/benchmark-reference.tex
In Fig. \ref{fig:benchmarking}, the \ac{DRIP} sequence is benchmarked against other methods, such as the \ac{SRIP} \cite{10061453}, the \ac{BnB} method in \cite{liu2018toward}, and the \ac{SQRBS} method introduced in \cite{aldayel2016successive}.

%% file: actions/sum-rate-computation.tex
For the sum rate computation, we follow the same approach as that in \cite{10061453} (c.f. equation (61)). 

%% file: actions/crossover-eta.tex
This is because, for high $\eta$ and high $\epsilon$, as \ac{SRIP} \cite{10061453} does not have a radar SINR constraint, \ac{SRIP} focuses all remaining resources to minimize \ac{MUI}. However, because \ac{DRIP} is forced to invest resources on the radar beam, \ac{DRIP} has fewer resources left to minimize interference for communication users. So, it sacrifices communication performance compared to \ac{SRIP}.
 For low $\eta$ and high $\epsilon$; because \ac{SRIP} \cite{10061453} only optimizes over temporal domain, the \ac{SRIP} sequence  prioritizes the \ac{PAPR} requirements, hence leaving few resources for \ac{MUI} minimization. \ac{DRIP}, on the other hand, can rely on the spatial domain to provide additional gains for users nearby radar beams.

%% file: actions/epsilon-eta-papr.tex
we analyze the behavior of the received constellations under varying PAPR regimes. It is crucial to clarify that the constellations plotted in Fig. \ref{fig:constellation}  represent the received signal at the users, denoted as $\pmb{HX} \approx \pmb{\mathbf{S}}$, rather than the transmit waveform. The distortion observed in these constellations, often showing spread, can be interpreted as \emph{implicit digital pre-distortion} induced by the optimization constraints. Specifically, as $\epsilon$ decreases, the transmit waveform is forced to adhere strictly to a deterministic radar chirp that carries no information, which in turn creates an unavoidable conflict with the random data symbols in $\pmb{S}$, resulting in residual \ac{MUI} that perturbs the received symbols away from their ideal positions. At high $\epsilon$ (e.g. \textcolor{black}{$\epsilon \geq 1.55$}), we can see the constellation resembles a clear \ac{QPSK} (in Fig. \ref{fig:constellation-01}) or 16-PSK shape (in Fig. \ref{fig:constellation-02}). This is because more priority is imposed over communications. On the other hand, as $\epsilon$ shrinks, the \emph{implicit digital pre-distortion} dominates within $\pmb{HX}$.

Besides, the nature of distortion is also heavily dependent on the \ac{PAPR} constraint. 
\textcolor{black}{\input{actions/revised-sentence-cloud}}
In contrast, when the PAPR constraint is tightened ($\eta \to 1 \dB$), the transmit waveform is constrained to a constant-modulus manifold. 
Consequently, the received constellations $\pmb{HX}$ in Fig. \ref{fig:constellation-03} and Fig. \ref{fig:constellation-04} introduce additional residual \ac{MUI} as $\pmb{HX}$ diverges away from $\pmb{S}$. This indicates that both \ac{PAPR} and similarity constraints attempt to distort the received constellation $\pmb{HX}$ appearing as spreads around the symbol's nominal position.

%% file: actions/revised-sentence-cloud.tex
As observed in Fig. \ref{fig:constellation-03} and Fig. \ref{fig:constellation-04}, when the \ac{PAPR} constraint is relaxed (e.g. $\eta = 4.5 \dB$), we observe that larger $\eta$ allows more amplitude variation in \pmb{X}, which helps $\pmb{HX}$ approaches the desired constellation $\pmb{S}$. For instance, in Fig. \ref{fig:constellation-04}, $\pmb{HX}$ becomes closer to the \ac{PSK} constellation for increasing $\eta$.

%% file: actions/gap-10-30.tex
Because the reference signal $\pmb{X}_0$ is constructed by repeating an \ac{LFM} chirp across all transmit antennas, this corresponds to a spatial beam focused at broadside $0^\circ$ by default. 
Therefore, the similarity constraint biases the optimization towards the reference angle of $0^\circ$. The steering onto $30^\circ$ requires a larger Euclidean deviation away from $\mathbf{X}_0$ than steering towards $10^\circ$, which makes the $10^\circ$ steering easier to satisfy in terms of the available similarity budget.
Hence, the performance gap is due to the geometric cost of steering the beam away from the reference waveform $\mathbf{X}_0$ (set at $0^\circ$). 
The hypothesis is further supported by the trend observed in Fig. \ref{fig:SINR-eps}, where the performance gap between the $10^\circ$ and $30^\circ$ targets further decreases as $\epsilon$ increases. 
At low $\epsilon$, the similarity budget is tight, which makes the higher steering cost of the $30^\circ$ target a bottleneck, thus reflected in lower \ac{SINR}.
At high $\epsilon$, the cost difference between steering to $10^\circ$ versus $30^\circ$ becomes more feasible to the available spatial resources, which allows the optimization to satisfy both targets equally.

%% file: actions/DRIP-compare-SINR.tex
In Fig. \ref{fig:RadarSINR_Benchmark}, we plot the average \ac{SINR} versus $\epsilon$ for a given \ac{SINR} requirement.
As compared to Fig. \ref{fig:SINR-eps}, where we see that \ac{DRIP} consistently achieves $48 \dB$ \ac{SINR} for high $\epsilon$ values. In contrast, all competing benchmarks, i.e. ADMM-SRIP \cite{10061453}, BnB \cite{liu2018toward}, and \ac{SQRBS} \cite{aldayel2016successive}, saturate at lower SINR levels, fluctuating between $-10\dB$ up to $30 \dB$. 
The $18 \dB$ gap validates the space-time design philosophy of \ac{DRIP}. While the benchmark methods primarily focus on optimizing the waveform's temporal properties (which transmit energy omni-directionally or with fixed gain), \ac{DRIP} exploits spatial degrees-of-freedom to beamform energy towards targets and nullify interference, which confirms its ability to satisfy stringent spatial constraints regardless of the temporal similarity requirements. 
In conclusion, \ac{DRIP} can maintain a high \ac{SINR} which outperforms baselines regardless of \ac{PAPR} constraint $\eta$.

%% file: actions/delay-doppler-angular.tex
It is worth noting that the similarity constraint forces the generated waveform to resemble a reference radar chirp. On the other hand, to achieve high radar \ac{SINR} and sharp angular peaks, the algorithm needs to manipulate the amplitude/phase variations of the \ac{DRIP} sequence across the transmit antennas. A conflict can arise because high similarity constrains the sequence into a specific temporal structure, which in turn can consume the degrees-of-freedom that would otherwise be used for spatial beamforming. 
Fig. \ref{fig:SINR-eps} along with Fig. \ref{fig:beampattern} depict such tradeoffs, whereby lower $\epsilon$ favors waveforms good for delay-Doppler tasks but lose beampattern accuracy, whereas higher $\epsilon$ can lead to poor delay-Doppler performance, but excellent beampatterns.

%% file: actions/eta-epsilon-papr-2.tex
If we relax the similarity constraint ($\epsilon \approx 2$) allows the optimization to fully exploit available resources, which results in waveforms that saturate the specified PAPR limits. Conversely, tightening $\epsilon$ forces the waveform to converge toward the  \ac{LFM} chirp, which suppresses the PAPR to near-unity regardless of the allowable threshold.

%% file: actions/eta-papr-mui.tex
In order to allude the trade-offs of the contradicting nature of \ac{MUI} and \ac{PAPR}, we have plotted Fig. \ref{fig:inst-MUI-PAPR} which plots the instantaneous \ac{MUI} vs instantaneous \ac{PAPR} for each Monte Carlo trial. The plot shows that a higher \ac{PAPR} leads to a lower achievable \ac{MUI}. For instance, when $\eta$ is set to as low as $0\dB$ (achieving \ac{PAPR}s close to $0\dB$, then the minimum achieved \ac{MUI} is about $10\dB$. On the other hand, when $\eta$ is relaxed to $3 \dB$, which achieves a \ac{PAPR} of exactly $3\dB$, then \ac{MUI} can go below $0\dB$. The simulation can guide the selection of threshold $\eta$ for PAPR in \eqref{eq:problem2}.

%% file: actions/average-similarity.tex
The average similarity is measured as the average of $\Vert \pmb{x} - \pmb{x}_0 \Vert^2$ taken over multiple Monte Carlo trials.

%% file: actions/HPA-impact.tex
\paragraph{\ac{HPA} impact}
To evaluate the impact of non-linear distortion, simulations were conducted as shown in Fig. \ref{fig:HPA}, where the sequence $\pmb{X}$ generated by \ac{DRIP} is subjected to \ac{HPA} clipping. The results demonstrate that the \ac{DRIP} performance maintains parity with the unclipped baseline (ideal linear amplification) provided that the scaling parameter $\eta$ remains below the input back-off clipping threshold. Specifically, for $\epsilon=2$ and an \ac{HPA} clipping level of $2\dB$, the average per-user rate coincides with the unclipped bound when $\eta < 2 \dB$. However, performance diverges for $\eta > 2 \dB$ due to saturation-induced distortion. Quantitatively, at an operating point of $\eta=3\dB$, the unclipped baseline achieves $6\bpspHz$, whereas a $2\dB$ clipping threshold reduces the rate to $5.4\bpspHz$. In this saturation regime, further reducing the clipping level results in a spectral efficiency penalty of approximately $0.7\bpspHz$ per $1\dB$ reduction.

%% file: actions/ZF-vs-MMSE.tex
\paragraph{\ac{ZF} vs \ac{MMSE}}
In Fig. \ref{fig:MMSE-vs-ZF}, we plot the per-user rate vs $\frac{1}{\sigma_v^2}$ for different values of $\eta$. While both equalizers reach the same peak performance for different $\eta$ values, the \ac{ZF} equalizer approaches the maximum rate faster, at lower $\frac{1}{\sigma_v^2}$ values, than the \ac{MMSE} equalizer because at high SNR, noise becomes negligible, where the \ac{MMSE} solution converges toward the ZF solution, which focuses purely on \ac{MUI} cancellation, rather than balancing the noise as well. Also, for a given target rate, say $2\bpspHz$, there is a $2.5\dB$ margin in terms of $\frac{1}{\sigma_v^2}$ in favor of \ac{ZF} as opposed to \ac{MMSE}.

%% file: sections/conclusions.tex
In this paper, we introduced \ac{DRIP}, which is a family of space-time \ac{ISAC} \textcolor{black}{sequences} capable of \ac{PAPR} adjustment.
The proposed space-time \ac{ISAC} \textcolor{black}{sequences}, i.e. \ac{DRIP}, abide by given \ac{PAPR} levels, while remaining similar to a given radar chirp, maintaining good beampattern properties towards desired directions, and rejecting interfering ones for multi-target sensing applications. 
As for communications, the proposed \ac{DRIP} waveforms intend to minimize \ac{MUI} for any given constellation.
As the optimization framework intended to generate such space-time \ac{ISAC} \textcolor{black}{sequences} is challenging to solve, we propose a \ac{BCCD} method that iterates between the different variables of the problem, which is guaranteed to converge to a waveform. 
Finally, simulation results demonstrate the superior performance, versatility, as well as \ac{ISAC} trade-offs of the proposed family of space-time \ac{ISAC} \textcolor{black}{sequences}.

%% file: sections/acks.tex
\textcolor{black}{This work is supported by Tamkeen under the Research Institute NYUAD grant CG017.}

%% file: sections/proof-qcqp.tex
In this part of the appendix, we prove that the problem in \eqref{eq:problem3-x} is \ac{QCQP}. 
Next, we translate each term appearing in \eqref{eq:problem3-x} to match the form of $(\mathcal{S}_{\rm{QCQP}})$ given in  \eqref{eq:problem-qcqp}.
The complex-valued variable in \eqref{eq:problem3-x} corresponds to $\pmb{x}_r$ as $\pmb{x}_r = \phi(\pmb{x})$. 
The cost in \eqref{eq:problem3-x} is quadratic in $\pmb{x}_r$ and hence we can pick
$\pmb{P}_{0} = \pmb{I}$ and 
$\pmb{q}_{0} = -2\phi(\pmb{x}_{\comm})$ to match the non-constant terms of the cost in \eqref{eq:problem-qcqp}.
The norm-equality constraint in \eqref{eq:problem3-x}, i.e. $\Vert \pmb{x} \Vert^2 = 1$, can be replaced by two inequality constraints, i.e. $\Vert \pmb{x}_r \Vert^2 \leq 1 $ and $\Vert \pmb{x}_r \Vert^2 \geq 1 $. Hence, choosing $\pmb{P}_1 = \pmb{I}$, $\pmb{P}_2 = - \pmb{I}$, $\pmb{q}_1 = \pmb{q}_2 = \pmb{0}$, and $r_1 = 1$ and $r_2 = -1$ account for constraint $\Vert \pmb{x} \Vert^2 = 1$.
The sphere constraint in \eqref{eq:problem3-x} can be represented by picking
$\pmb{P}_{3} = \pmb{I}$,
$\pmb{q}_{3} = -2\phi(\pmb{x}_0)$, and
$r_{3} =  \Vert \pmb{x}_0 \Vert^2 - \epsilon^2$.
The quadratic-inequality constraint in \eqref{eq:problem3-x}, i.e. $\pmb{x}^H \pmb{F}_p \pmb{x} \leq \frac{\eta}{N_TL}$, can be realized by picking $\pmb{P}_{p+3} = \phi(\pmb{F}_p)$, $\pmb{q}_{p+3} = \pmb{0}$ and $r_{p+3} = \frac{\eta}{N_TL}$, for all $p= 1 \ldots N_T L$.
Last but not least, the radar \ac{SINR} constraints in \eqref{eq:problem3-x} can also be arranged via \ac{QCQP} constraints. In realizing so, we first interchange the role of $\widehat{g}_q$ and $\pmb{x}$. For that, we have
$\widehat{g}_q = \frac{{\pmb{x}}^H \pmb{S}_{1,q} {\pmb{x}}}{{\pmb{x}}^H \pmb{S}_{2,q} {\pmb{x}} + \sigma_r^2 \Vert \widehat{\pmb{u}}_q \Vert^2}$, $\forall q = 1 \ldots Q$, where 
\begin{align}
\label{eq:S1q}
	\pmb{S}_{1,q} 
	&=
	\sigma_q^2
	\big( \pmb{I}_L \otimes \pmb{\Pi}(\theta_q) \big)^H
	\widehat{\pmb{u}}_q
	\widehat{\pmb{u}}_q^H
	\big( \pmb{I}_L \otimes \pmb{\Pi}(\theta_q) \big) \\
	\begin{split}
	\pmb{S}_{2,q} 
	&=
	\sum\nolimits_{q' \neq q}
	\sigma_{q'}^2
	\big( \pmb{I}_L \otimes \pmb{\Pi}(\theta_{q'}) \big)^H
	\widehat{\pmb{u}}_q
	\widehat{\pmb{u}}_q^H
	\big( \pmb{I}_L \otimes \pmb{\Pi}(\theta_{q'}) \big) 
			 \\
	&+
	\sum\nolimits_{i}
	\bar{\sigma}_{i}^2
	\big( \pmb{I}_L \otimes \pmb{\Pi}(\bar{\theta}_{i}) \big)^H
	\widehat{\pmb{u}}_q
	\widehat{\pmb{u}}_q^H
	\big( \pmb{I}_L \otimes \pmb{\Pi}(\bar{\theta}_{i}) \big) \label{eq:S2q}
	\end{split}
\end{align}
Therefore, choosing
$\pmb{P}_{N_TL + 3 +  q} =\phi(\bar{g}_q \pmb{S}_{2,q} - \pmb{S}_{1,q}) $,
$\pmb{q}_{N_TL + 3 +  q} = \pmb{0}$, and
$r_{N_TL + 3 +  q} =-  \bar{g}_q  \Vert  \widehat{\pmb{u}}_q  \Vert^2$ for all $q = 1 \ldots Q$ finalizes the proof.

%% file: sections/proof-converge-stationary.tex
We first take the gradient of \eqref{eq:L-rho-hat} with respect to $\pmb{x}_r$ as 
\begin{equation*}
\begin{split}
	\nabla_{\pmb{x}_r}\widehat{\mathcal{L}}_\rho(\pmb{x}_r,\pmb{\lambda})
	&=
	2 \pmb{P}_0 \pmb{x}_r 
	+  \pmb{q}_0 
	\\ &+
	\rho
	\sum\nolimits_i 
	\left[\frac{\lambda_i}{\rho}
	+\pmb{x}_r^T \pmb{P}_i \pmb{x}_r
	+ \pmb{q}_i^T \pmb{x}_r
	- r_i\right]^+\pmb{g}_i.
\end{split}
\end{equation*}
where $\pmb{g}_i = 2\pmb{P}_i \pmb{x}_r
	+ \pmb{q}_i$ if $\frac{\lambda_i}{\rho}
	+\pmb{x}_r^T \pmb{P}_i \pmb{x}_r
	+ \pmb{q}_i^T \pmb{x}_r
	- r_i \geq 0$, else $\pmb{g}_i = \pmb{0}$, which implies that the above gradient can be expressed as
\begin{equation}
\label{eq:grad-of-hat-L-rho}
\begin{split}
	&\nabla_{\pmb{x}_r}\widehat{\mathcal{L}}_\rho(\pmb{x}_r,\pmb{\lambda})
	=
	2 \pmb{P}_0 \pmb{x}_r 
	+  \pmb{q}_0 
	\\ &+
	\rho
	\sum\nolimits_i 
	\left[\frac{\lambda_i}{\rho}
	+\pmb{x}_r^T \pmb{P}_i \pmb{x}_r
	+ \pmb{q}_i^T \pmb{x}_r
	- r_i\right]^+(2\pmb{P}_i \pmb{x}_r
	+ \pmb{q}_i).
\end{split}
\end{equation}	
But notice that when the gradient in \eqref{eq:grad-of-hat-L-rho} is evaluated at $(\pmb{x}_r^{(n)},\pmb{\lambda}^{(n-1)})$, and with the aid of \eqref{eq:lambda_i_n}, it follows that
\begin{equation}
\label{eq:grad-of-hat-L-rho-2}
	\nabla_{\pmb{x}_r}\widehat{\mathcal{L}}_\rho^{(n)} 
	=
	2 \pmb{P}_0 \pmb{x}_r^{(n)}
	+  \pmb{q}_0 
	+
	\sum\nolimits_i 
	\lambda_i^{(n)}(2\pmb{P}_i \pmb{x}_r^{(n)}
	+ \pmb{q}_i),
\end{equation}	
where $\nabla_{\pmb{x}_r}\widehat{\mathcal{L}}_\rho^{(n)} \triangleq \nabla_{\pmb{x}_r}\widehat{\mathcal{L}}_\rho(\pmb{x}_r^{(n)},\pmb{\lambda}^{(n-1)})$; which means that $\pmb{x}_r^{(n+1)}$ minimizes $\pmb{x}_r^T \pmb{P}_0 \pmb{x}_r
	+  \pmb{q}_0^T\pmb{x}_r 
	+
	\sum\nolimits_i 
	\lambda_i^{(n)}(\pmb{x}_r^T \pmb{P}_i \pmb{x}_r
	+ \pmb{q}_i^T \pmb{x}_r)$.
Denoting 
\begin{equation*}
	\Delta h^{(n)} = (\pmb{x}_r^{(n+1)})^T\pmb{P}_0 \pmb{x}_r^{(n+1)}
	+  \pmb{q}_0^T\pmb{x}_r^{(n+1)} - \bar{\pmb{x}}_r^T \pmb{P}_0 \bar{\pmb{x}}_r
	+  \pmb{q}_0^T\bar{\pmb{x}}_r,
\end{equation*} 
we have that
\begin{equation*}
\begin{split}
\Delta h^{(n)}
	&\leq 
	\sum\nolimits_i
	\lambda_i^{(n)}
	\left(
	\bar{\pmb{x}}_r^T \pmb{P}_i \bar{\pmb{x}}_r
	-
	(\pmb{x}_r^{(n+1)})^T \pmb{P}_i \pmb{x}_r^{(n+1)}
	\right) \\ &+
	\sum\nolimits_i
	\lambda_i^{(n)}
	\left(
	 \pmb{q}_i^T \bar{\pmb{x}}_r
	- \pmb{q}_i^T \pmb{x}_r^{(n+1)}
	\right).
\end{split}
\end{equation*}
Therefore, as 
$\lim\nolimits_{n \rightarrow \infty} (\pmb{x}_r^{(n+1)})^T \pmb{P}_i \pmb{x}_r^{(n+1)} = \bar{\pmb{x}}_r^T \pmb{P}_i \bar{\pmb{x}}_r$ (namely $\left| (\pmb{x}_r^{(n+1)})^T \pmb{P}_i \pmb{x}_r^{(n+1)} - \bar{\pmb{x}}_r^T \pmb{P}_i \bar{\pmb{x}}_r \right| \leq \epsilon_1 $ for every $\epsilon_1 > 0$ and a given iteration $N$ for all $n \geq N$)
and 
$\lim\nolimits_{n \rightarrow \infty} \pmb{q}_i^T \pmb{x}_r^{(n+1)}= \pmb{q}_i^T \bar{\pmb{x}}_r$ (namely $\left| \pmb{q}_i^T \pmb{x}_r^{(n+1)}- \pmb{q}_i^T \bar{\pmb{x}}_r \right| \leq \epsilon_2$ for every $\epsilon_2 > 0$ and a given iteration $N$ for all $n \geq N$), we get $\lim\nolimits_{n \rightarrow \infty} \Delta h^{(n)} = 0$, i.e. $\lim\nolimits_{n \rightarrow \infty} \Delta h^{(n)} = 0$ and the algorithm terminates, which finalizes the proof.

%% file: sections/outer-convergence.tex
At iteration number $k$, the space-time \ac{ISAC} \textcolor{black}{sequence} and beamforming vectors are assumed to satisfy the \ac{DRIP} constraints, i.e. the constraints of optimization problem $(\widetilde{\mathcal{P}}_{\rm{DRIP}})$ in \eqref{eq:problem3}, namely 
\begin{equation}
\begin{aligned}
\begin{cases}
 &  \textcolor{black}{\Vert \pmb{x}^{(k)} \Vert^2 = 1,} \\ 
 & (\pmb{x}^{(k)})^H \pmb{F}_p \pmb{x}^{(k)} \leq \frac{\eta}{N_TL} , \quad \forall p \\
 & \pmb{x}^{(k)} \in \mathcal{B}_{\epsilon}(\pmb{x}_0), \quad  g_q(\pmb{x}^{(k)},\widehat{\pmb{u}}_q^{(k)}) \geq \bar{g}_q, \quad \forall q.
\end{cases}
\end{aligned}
\end{equation} 
In the following iteration, $\pmb{x}^{(k+1)}$ is obtained by solving
\begin{equation}
\label{eq:problem-at-kp1}
\begin{aligned}
\begin{cases}
\min\nolimits_{ \pmb{x} }&  \Vert  \pmb{x} - \pmb{x}_{\comm} \Vert^2 \\
\textrm{s.t.}
 &  \textcolor{black}{\Vert \pmb{x} \Vert^2 = 1,} \\ 
 & \pmb{x}^H \pmb{F}_p \pmb{x} \leq \frac{\eta}{N_TL} , \quad \forall p \\
 & \pmb{x} \in \mathcal{B}_{\epsilon}(\pmb{x}_0), \quad g_q(\pmb{x},\widehat{\pmb{u}}_q^{(k+1)}) \geq \bar{g}_q, \quad \forall q,
\end{cases}
\end{aligned}
\end{equation}
where $\widehat{\pmb{u}}_q^{(k+1)}$ solves \eqref{eq:problemu}, given $\pmb{x}^{(k)}$. It follows that
\begin{equation}
	g_q(\pmb{x}^{(k)},\widehat{\pmb{u}}_q^{(k+1)})
	\geq 
	g_q(\pmb{x}^{(k)},\widehat{\pmb{u}}_q^{(k)})
	\geq 
	\bar{g}_q, \quad \forall q.
\end{equation}
In turn, since $( \pmb{x}^{(k+1)}$,$\lbrace \widehat{\pmb{u}}_q^{(k+1)} \rbrace_{q=1}^q )$ solves \eqref{eq:problem-at-kp1}, then it should satisfy its constraints and hence is a feasible solution.
In addition,
\begin{equation*}
	\Vert  \pmb{x}^{(k+1)} - \pmb{x}_{\comm} \Vert^2
	\leq
	\Vert  \pmb{x}^{(k)} - \pmb{x}_{\comm} \Vert^2.
\end{equation*}
By induction, the algorithm converges and the proof is done.